\documentclass[fleqn,usenatbib]{mnras}

\usepackage{newtxtext,newtxmath}

\usepackage[T1]{fontenc}
\usepackage{ae,aecompl}

\usepackage{graphicx}	
\usepackage{amsmath}	
\usepackage{amssymb}	
\usepackage{mwe}
\usepackage{color}

\usepackage{etoolbox}
\makeatletter
\patchcmd\@combinedblfloats{\box\@outputbox}{\unvbox\@outputbox}{}{\errmessage{\noexpand patch failed}}
\makeatother

\title[Planet gap opening across stellar masses]{Planet gap opening across stellar masses}

\author[C. A. Sinclair et al]{
Catriona A. Sinclair,$^{1}$\thanks{E-mail: cas213@cam.ac.uk}
Giovanni P. Rosotti,$^{1,2}$
Attila Juhasz,$^{1}$
Cathie J. Clarke$^{1}$
\\
$^{1}$Institute of Astronomy, Madingley Road, Cambridge CB3 OHA, UK\\
$^{2}$Leiden Observatory, Leiden University, PO Box 9513, NL-2300 RA Leiden, the Netherlands
}

\date{Accepted XXX. Received YYY; in original form ZZZ}

\pubyear{2019}

\begin{document}
\label{firstpage}
\pagerange{\pageref{firstpage}--\pageref{lastpage}}
\maketitle

\begin{abstract}
Annular structures in proto-planetary discs, such as gaps and rings, are now ubiquitously found by high-resolution ALMA observations. Under the hypothesis that they are opened by planets, in this paper we investigate how the minimum planet mass needed to open a gap varies across different stellar host masses and distances from the star. The dependence on the stellar host mass is particularly interesting because, at least in principle, gap opening around low mass stars should be possible for lower mass planets, giving us a look into the young, low mass planet population. Using dusty hydrodynamical simulations, we find however the opposite behaviour, as a result of the fact that discs around low mass stars are geometrically thicker: gap opening around low mass stars can require more massive planets. Depending on the theoretical isochrone employed to predict the relationship between stellar mass and luminosity, the gap opening planet mass could also be independent of stellar mass, but in no case we find that gap opening becomes easier around low mass stars. This would lead to the expectation of a {\it lower} incidence of such structures in lower mass stars, since exoplanet surveys show that low mass stars have a lower fraction of giant planets. More generally, our study enables future imaging observations as a function of stellar mass to be interpreted using information on the mass vs. luminosity relations of the observed samples. 
\end{abstract}


\section{Introduction}\label{sec:intro}
With the Atacama Large Millimetre Array (ALMA) telescope having reached full operation, the field of proto-planetary discs is undergoing a rapid observational expansion. Thanks to the order of magnitude improvement in spatial resolution, we now have the possibility of resolving the signatures of planets in formation in these discs, in this way transforming planet formation into an observational field.

The most striking result of these observations is the ubiquity of annular structures, colloquially described as gaps and rings. While some discs do show alternative structures like spirals or crescents \citep[e.g.,][]{2013Sci...340.1199V,2015ApJ...812..126C,2016Sci...353.1519P,2018ApJ...853..162B,2018A&A...619A.161C}, most of the discs observed at high resolution are characterised by axisymmetric structures. The prevalence of axisymmetric structures was already clear from the publication of several high-resolution observations targeting individual sources \citep{Brogan2015,2016PhRvL.117y1101I,2016ApJ...820L..40A,2017A&A...597A..32V,2017A&A...600A..72F,2017ApJ...840...23L,2018A&A...610A..24F,2018MNRAS.475.5296D,2018ApJ...866L...6C} and from the survey in Taurus \citep{2018ApJ...869...17L}, but recently it was made even clearer by the publication of DSHARP \citep{2018ApJ...869L..41A,2018ApJ...869L..42H}, a homogeneous high resolution survey of 20 discs. All the 18 single disc systems show annular structure; only a minority also exhibit additional structure superimposed on the background annular structure, with 3 showing spirals and 2 showing crescents.

Planets naturally create annular structures in discs \citep[e.g.,][]{2004A&A...425L...9P,2012A&A...545A..81P,Picogna2015,Rosotti2016,2017MNRAS.469.1932D,2018ApJ...869L..47Z} and are therefore the leading explanation for these structures. There are however also alternative interpretations. An intriguing idea is that these observed structures do not correspond to real features in the disc surface density, but they are caused by opacity changes \citep{2015ApJ...806L...7Z,2016ApJ...821...82O,2017ApJ...845...68P,2017A&A...600A.140S}. In this view the change in opacity should happen at the locations of snowlines, where the most abundant molecules change from the solid to the gas phase, triggering compositional changes in the dust. Recent work \citep{2018ApJ...869L..42H} however has put this idea into question since the location of most of the observed gaps do not correspond to the predicted location of the snowlines. In addition, some gaps do correspond to physical structures in the gas surface density, as shown by gas emission line profiles \citep{2016PhRvL.117y1101I,2017A&A...600A..72F} and kinematically derived rotation curves \citep{2018ApJ...860L..12T,2018ApJ...868..113T}. Some gaps \citep{2018ApJ...869L..48G} are so deep that they cannot be accounted for by opacity variations, and they must correspond to a real depletion in surface density. Still open instead is the possibility that these structures are created by the interplay between magnetohydrodynamics (MHD) and dust dynamics  \citep[e.g.,][]{2015A&A...574A..68F,2018A&A...609A..50D}. In a similar way to planets, MHD processes can also alter the gas surface density and its pressure profile, causing a variation in the dust radial velocity and therefore its surface density. This possibility has received less attention than the planet hypothesis and it is currently less clear how to distinguish between the two.

Rather than contributing directly to this debate, in this paper we will focus on the planet hypothesis and explore its consequences. One of the fundamental questions in this case is what is the range of masses of the putative planets. Broadly speaking, on the upper end of the planet mass range we can exclude in most cases that these planets are gas giants of several Jupiter masses. These planets tend to create non-axisymmetric structures like spiral arms and crescents (see for example the gallery of simulated observations in \citealt{2018ApJ...869L..47Z}) and allow for very little passage of dust through the planet orbit \citep{2012ApJ...755....6Z}, depleting most of the disc interior to the planet location. As such, they are more commonly invoked to explain the few discs that show prominent spiral arms \citep{2015MNRAS.451.1147J,2016ApJ...816L..12D,2017ApJ...839L..24M,2018MNRAS.474L..32J} or the so-called ``transition discs'' \citep[e.g.,][]{2015A&A...580A.105P,2016MNRAS.459L..85D}, rather than the gapped discs (though with the recent high-resolution observations the distinction between the two categories is becoming blurred). On the lower end of the mass range instead, it is well known \citep{Lin1993,Crida2006} that there is a minimum threshold mass needed to open a gap. In the \textit{gas} case, it is very well known that this depends on the disc aspect ratio and viscosity; in general, gap opening requires higher-mass planets if the disc is thicker (i.e., hotter) and more viscous. ALMA observations however probe the \textit{dust}. Because it is easier to open gaps in dust than in gas, this mass threshold is \textit{quantitatively} different (lower) than for the conventional gas case. Ultimately, however, the mass threshold should not change \textit{qualitatively} since the the dust morphology is set by the underlying gas profile. In particular, the gas radial pressure gradient determines the dust radial drift velocity, and in turn its surface density \citep{2012ApJ...755....6Z,Rosotti2016}.

While the dependence of the gap-opening mass on the disc parameters has been extensively studied in previous works, a less explored aspect is the dependence on the stellar mass. Since the gaps are of dynamical origin, the real underlying parameter is the planet-star mass ratio, not the absolute planet mass. Targeting lower mass stars could then open the exciting possibility of detecting planets significantly lower in mass. Since the threshold mass around a solar mass star is typically in the super-Earth regime \citep{Rosotti2016}, in principle the sensitivity around an ultracool dwarf should be comparable to an Earth mass. Conceptually, this is a similar motivation to the study of exoplanets around mature low mass stars with the conventional techniques of transits and radial velocity, that has led, for example, to the discovery of the TRAPPIST1 system \citep{2016Natur.533..221G}. However, the dependence on the aspect ratio mentioned above also needs to be taken into account. It is well known \citep[e.g.,][]{2012A&A...539A...9M,2013A&A...554A..95P} that discs around low mass stars and brown dwarves are geometrically thicker than those around solar mass stars, as a consequence of the reduced gravitational potential\footnote{In general, this is is more important than the fact that these discs are colder due to the fainter central star. We will discuss this in the detail in the rest of the paper.}. This effect makes it \textit{harder} to open gaps around lower mass stars, in the opposite direction to what we have described before. 

Determining which of the two effects is dominant is the purpose of this paper. Building on the disc-planet interaction dusty simulations presented by \citet{Rosotti2016}, in this work we will explore the dependence of the gap-opening mass on the stellar mass. This paper is observationally focused and our definition of gap opening is therefore an observable gap \textit{in the dust}. A key aspect of this work is that we compute the disc temperature rather than leaving it as a free parameter (as commonly done in hydrodynamic simulations). 

This paper is structured as follows. We explain our methodology in section \ref{sec:methods} and present our results in section \ref{sec:results}. We then discuss the implications of our results for observations targeting low mass stars and for the planet hypothesys for the origin of gaps in section \ref{sec:discussion} and finally draw our conclusions in section \ref{sec:conclusions}.

\section{Methods}\label{sec:methods}
Our methodology consists of running hydrodynamical simulations of the gas and dust dynamics with the code FARGO to study how these components respond to the presence of a planet. 
We then use the radiative transfer code RADMC-$3$D\footnote{\url{http://www.ita.uni-heidelberg.de/dullemond/software.radmc-3d/}} to compute the disc temperature and generate synthetic images. 
Finally, we use the CASA tool to simulate realistic ALMA observations. 
We detail this work flow in the following sections \ref{ssec:hydrosim}, \ref{ssec:RT} and \ref{ssec:CASA}. 

One particular aspect to note is that for the hydrodynamical simulations the disc temperature, typically parameterised through the disc aspect ratio $h/r$, is a free parameter. 
This is particularly important because the aspect ratio has a major impact on the minimum gap opening planet mass (MGOPM). 
To run realistic hydrodynamical simulations, it is thus necessary to know how the disc aspect ratio varies as a function of stellar mass and disc radius. 
To this end, we perform a preliminary set of calculations with RADMC-$3$D, not containing any planet, that we describe in section \ref{sssec:dischr}.

\subsection{Hydrodynamical Simulations}\label{ssec:hydrosim}
The simulations we present in this paper use a custom version of the \textsc{fargo-3d} code \citep{2016ApJS..223...11B}, modified to include dust dynamics as described in \citet{Rosotti2016}; we refer the reader to that paper for more details on the dust algorithm. Briefly, dust is described as a pressure-less fluid, evolving because of gravity, gas drag and diffusion. We implement gas drag using a semi-implicit algorithm that automatically reduces to the short-friction time approximation for tightly coupled dust (so that the timestep does not become vanishingly small) and to an explicit update for loosely coupled grains. For the diffusion, we use a diffusion coefficient equal to the shear viscosity coefficient of the gas (in other words, the Schmidt number is 1).

We use 2D cylindrical coordinates and dimensionless units in which the orbital radius of the planet ($r_p$) is at unity, the unit of mass is that of the central star and the unit of time is the inverse of the Kepler-frequency of the planet. The inner radial boundary of our grid is at 0.5 $r_p$ and the outer boundary at $3 r_p$; we use non reflecting boundary conditions at both boundaries. While \citet{Rosotti2016} fixed the dust density to its initial value at the inner boundary, here we allow the dust density to drop below this value (see also \citealt{Meru2018}). The resolution is 450 and 1024 uniformly spaced cells in the radial and azimuthal direction, respectively. The planet is kept on a circular orbit whose orbital parameters are not allowed to vary (see \citealt{Meru2018} for a study on the effects of migration). The surface density profile follows $\Sigma \propto r^{-1}$; since we fix the planet orbital parameters the value of the normalisation constant is arbitrary. Finally, in this paper we use the $\alpha$ prescription of \citet{ShakuraSunyaev} for what concerns the viscosity and we assume $\alpha = 10^{-3}$. 

The dynamics of the dust depends on the magnitude of the acceleration induced by gas drag, i.e. $a=\Delta v/t_s$, where $\Delta v$ is the relative velocity between the gas and the dust $t_s$ is the stopping time which depends on the grain properties and chiefly on the grain size. This is typically expressed in units of the local Keplerain time and called Stokes number $St=t_s/\Omega_K^{-1}$. To keep our simulations scale-free, each dust species in our simulation has a constant Stokes number. Once the scaling parameters of the disc have been chosen, it is then possible to convert the Stokes numbers to physical grain sizes as we explain in the next section. We use 5 dust sizes, with Stokes numbers (logarithmically spaced) ranging from $2 \times 10^{-3}$ to 0.2.

As mentioned before, in this paper the aspect ratio plays an important role. Therefore, we run a grid of models with aspect ratios of 0.025, 0.033, 0.04, 0.05, 0.066, 0.085 and 0.1 at the planet location; we discuss in section \ref{sssec:dischr} the link with the physical separation of the planet from the star. For any chosen normalisation, the aspect ratio in the disc varies in a power-law fashion with a flaring index of 0.25.

The goal of this paper is to study, for any planet location, the minimum gap opening planet mass. For this reason, for every different aspect ratio we run simulations with different planet masses, which for a star of 1 $M_\odot$ correspond to planet masses of 2.5, 4, 8, 12, 20, 60 and 120 $M_\oplus$. Note that we do not run the full range of planet masses for every aspect ratio, because in some cases it is already obvious that a planet of a given mass is able (or not) to open a gap. We then re-use the grid of simulations when considering different stellar masses, but note that, because in any simulation the planet-star mass ratio is fixed, this leads to different absolute planet masses.

\subsection{Radiative Transfer} \label{ssec:RT}
\paragraph*{Basic Disc Properties}\label{sssec:discprop}
In order to simulate images of discs around stars of different masses we construct disc models by adopting basic disc properties and scaling laws for the mass and radius of the disc with stellar mass. 
All discs were assumed to have a fixed gas to dust ratio of $100:1$. 
In these calculations we are interested in the disc temperature, which is mostly set by the small, well-coupled grains; therefore we do not take into account dust settling.
The unperturbed surface density profile of each disc was assumed to be azimuthally symmetric and inversely proportional to radius:
\begin{equation} \label{eqn:sigma}
	\Sigma(r, \phi) = \Sigma(r) \propto \frac{1}{r},
\end{equation}
the same as the relationship adopted by the hydrodynamical simulations discussed in section \ref{ssec:hydrosim}

\subsubsection{Scaling Relations}\label{sssec:scalings}  
The scaling relation between protoplanetary disc mass ($M_{d}$) and stellar mass ($M_{*}$) is well constrained by observations:
\begin{equation}
	\label{eqn:Md_M*}
    M_{d} \, \propto \, M_{*}^{1.3}.
\end{equation}

These observations (of the Chamaeleon I star forming region) correct for the variation in stellar luminosity, and therefore disc temperature, with stellar mass \citep{Pascucci2016}. 

Several power laws describing the scaling relationship between protoplanetary disc radius ($R_{d}$) and disc mass have been proposed (for example, by \cite{Tazzari2017} and \cite{Tripathi2017}, among others). 
For this work we adopt the scaling derived from temperature corrected data \citep{Andrews2018}. 
Combined with equation \ref{eqn:Md_M*}, this gives:
\begin{equation}
    R_{d} \, \propto \, M_{*}^{0.6}.
\end{equation}

Typical values for the outer disc radii and disc mass for a $1 \, \mathrm{M_{\sun}}$ mass star were adopted as the constants of proportionality, giving the following scaling relations. 
\begin{equation}
\label{eqn:discmass_starmass}
    M_{d} = \bigg(\frac{M_{*}}{\mathrm{M_{\sun}}} \bigg)^{1.3} \, 0.01 \, \mathrm{M_{\sun}}
\end{equation}
\begin{equation}
\label{eqn:discradius_starmass}
    R_{d} = \bigg(\frac{M_{*}}{\mathrm{M_{\sun}}} \bigg)^{0.61} \, 100 \, \mathrm{AU}
\end{equation}

In reality, observations show some spread around these average values, but we neglect this to reduce the number of free parameters in our simulations. For what concerns the outer radius, this serves only as a guide to know how large the average disc (for a given stellar mass) is; the results we will present in the following sections contain the necessary information to know MGOPM at large radii in case the radius of an individual disc is larger than the average value. For what concerns the disc mass, it should come as a caveat that there are instead physical effects that we are neglecting: namely, the variation of the mid-plane disc temperature with the disc surface density (which scales as $\Sigma^{-1/4}$, see appendix \ref{sec:append_temp}) and the variation of the Stokes number with surface density (although the grain size might also depends on the disc surface density, e.g. \citealt{Birnstiel2012}).

\subsubsection{Pre-main Sequence Evolutionary Models}\label{sssec:stardiscparam}  
The properties of the central star, specifically the mass, radius and effective temperature, are required by the radiative transfer code. 
These properties were extracted from a series of pre-main sequence evolutionary tracks from \citet{Siess2000}, and are shown in table \ref{tab:stellarprop}. 
For all discs considered in this work the age of the system was assumed to be $10^{6}$ years. 
We comment on the choice of the pre-main sequence evolutionary tracks in section \ref{sec:robustness} and \ref{sec:spread}.

\begin{table}
    \centering
    \begin{tabular}{| c | c | c | c |}
        \hline
        $\mathbf{ M_{*} \, \big[ M_{\sun} \big] } $ & $\mathbf{ L_{*} \, \big[ L_{\sun} \big] } $ & $\mathbf{ R_{*} \, \big[ R_{\sun} \big] } $ & $\mathbf{T_{eff} \, \big[ K \big]}$ \\
        \hline
        $1.0$ & $2.33$ & $2.62$ & $4278$ \\
        \hline
        $0.7$ & $1.72$ & $2.54$ & $4024$ \\
        \hline
        $0.3$ & $0.69$ &  $2.32$ & $3360$ \\
        \hline 
    \end{tabular}
    \caption{The stellar properties (mass $\big( M_{*} \big) $, luminosity $ \big( L_{*} \big)$, radius $ \big( R_{*} \big)$, and effective temperature $ \big( T_{eff} \big)$) used in the radiative transfer code RADMC-$3$D, obtained from the pre-main sequence evolutionary models at an age of $10^{6} \, \mathrm{years}$. }
    \label{tab:stellarprop}
\end{table}

\subsubsection{Model Disc Aspect Ratio Profiles}\label{sssec:dischr}
\begin{figure} 	
	\centering
    \includegraphics[width = 0.4\textwidth]{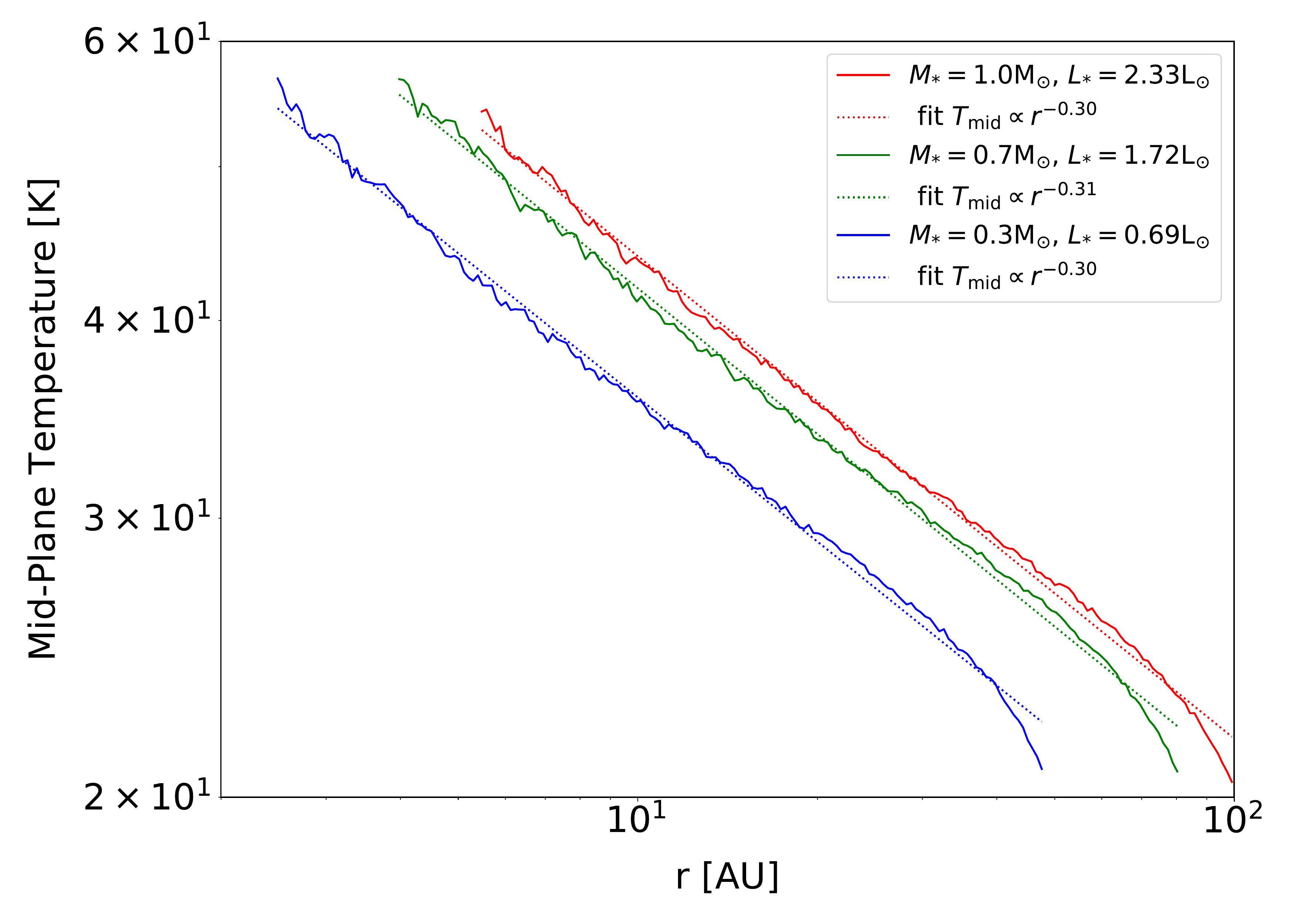}
    \caption{Mid-plane temperature profiles calculated for the three stellar masses using the RADMC$3$D \texttt{ppdisk} model, assuming the stellar parameters given in table \ref{tab:stellarprop}. 
    The profile produced by this model is shown as a solid line, and the power law fit to each profile is shown as a dashed line. }
	\label{fig:Temp_Scaling}
\end{figure}

The 3D temperature profiles of a series of systems comprising a central star and an empty (containing no planets) model disc were obtained using the default RADMC-$3$D model \verb+ppdisk+. 
Mid-plane temperature profiles were calculated for stellar masses of $M_{*} = 0.3$, $0.7$ and $1.0 \, \mathrm{M_{\sun}}$, with the relevant stellar properties (luminosity, radius and effective temperature) obtained from the pre-main sequence evolutionary models discussed in section \ref{sssec:stardiscparam}. 
A power law fit was applied to these temperature profiles and used to calculate the aspect ratio profiles for each system. 
These temperature profiles are shown in figure \ref{fig:Temp_Scaling}.

The aspect ratio is given by the ratio of pressure scale height, $h$, to radial position within the disc, $r$:
\begin{equation}
	\label{eqn:aspectratio}
    \frac{h}{r} = \frac{c_{s}}{v_{k}} = \sqrt{\frac{k_{B} T}{\mu m_{p}} \frac{r}{G M_{*}}},
\end{equation}
where $c_{s}$ is the isothermal sound speed, given by $c_{s} = \sqrt{\frac{k_{B} T}{\mu m_{p}}}$ where $T$ is the temperature in the disc and $\mu = 2.3$ is the mean molecular weight. 

The locations in each disc at which the aspect ratio is equal to the values for which hydrodynamical simulation data were extracted and are shown in table \ref{tab:aspectratio}. 
The aspect ratio at the location of the planet in the hydrodynamical simulation sets the semi-major axis of the planet in the radiative transfer model, and additionally sets the length scale of the data.  

The relationship between temperature at a given position within the disc and stellar mass was investigated using linear regression and found to scale approximately as:
\begin{equation}
\frac{T}{\mathrm{K}} \propto \bigg( \frac{M_{*}}{\mathrm{M_{\odot}}} \bigg)^{0.15}
\label{eqn:T_M}
\end{equation}
We emphasise that the above scaling only holds in the case of a specific assumption about the stellar mass luminosity relation; for the particular case used here, $L \propto M_\ast^{1.07}$ when measured between $0.3$ and $1.0 M_\ast$.
 
\begin{table*}
    \centering
    \begin{tabular}{| c | c | c | c | c | c | c | c |}
        \hline
        Semi-Major Axis [AU] & \multicolumn{7}{|c|}\textbf{{Aspect Ratio}}\\
        \hline
        \textbf{Stellar Mass} $\big[ \mathrm{\mathbf{M_{\sun}}} \big]$ &  $\mathbf{0.025}$  &  $\mathbf{0.033}$  &  $\mathbf{0.04}$  &  $\mathbf{0.05}$  &  $\mathbf{0.066}$  &  $\mathbf{0.085}$  &  $\mathbf{0.1}$   \\
        \hline
        $ \mathbf{0.3} $ & - & $1.1$* & $1.9$ & $3.6$ & $8.4$ & $17.9$ & $29.5$ \\
        \hline
        $ \mathbf{0.7} $ & $1.3$* & $2.9$ & $5.2$ & $10.1$ & $23.2$ & $50.3$ & - \\
        \hline
        $ \mathbf{1.0} $ & $2.5$ & $5.3$ & $9.7$ & $19.8$ & $36.2$ & $75.7$ & - \\
        \hline 
    \end{tabular}
    \caption{The semi-major axis (in $\mathrm{AU}$) corresponding to the aspect ratio of the hydrodynamical simulations, for the three different stellar masses considered in this project. Situations where the aspect ratio profile of the disc does not encompass the simulation aspect ratio are marked by a dash (-). 
    The values marked by asterisks correspond to semi-major axes for which no radiative transfer simulations were run, due to their extreme proximity to the inner edges of the discs in question. }
    \label{tab:aspectratio}
\end{table*}

\subsubsection{Conversion of Hydrodynamical Simulation Data} \label{sssec:conversion}

\paragraph*{Extrapolation}  
We use a $3$D spherical coordinate system with $N_{r} = 256$ logarithmically spaced points, $N_{\theta} = 100$ points distributed linearly in the three intervals $[0, \frac{\pi}{3}]$, $[\frac{\pi}{3}, \frac{2 \pi}{3}]$, $[\frac{2 \pi}{3}, \pi]$, as $N_{\theta} = \{ 10,80,10 \} $, and $N_{\phi} = 200$ grid points linearly spaced from $0$ to $2 \pi$ in the azimuthal direction. 
The first six radial cells of the hydrodynamical simulation data were excluded as they show artefacts caused by the inner boundary condition before the hydrodynamical simulation data is mapped onto this grid. 
In some cases the simulation data does not cover the entire extent of the model disc and in these situations the surface density was extrapolated out to the edges of the disc.

\paragraph*{Interpolation}
The dust population was modeled as ten logarithmically spaced grain size bins between $a = 10^{-5} \, \mathrm{cm}$ and $0.1 \, \mathrm{cm}$, with a size distribution described by:
\begin{equation}
    \label{eq:size_dist}
	\frac{\mathrm{d}N}{\mathrm{d}a} \propto a^{-3.5}.
\end{equation}

The opacities of the dust grain populations are calculated from the grain size and the mass absorption coefficients. 
The mass absorption coefficients used in this work were calculated using Mie theory, using the optical properties of astronomical silicates from \citet{Weingartner2001}. 

In order to compute the surface density of each dust species we compute the Stokes number, $St$, from the normalised gas surface density. Assuming that all particles are in the Epstein regime, the Stokes number obeys:
\begin{equation}
	\label{eq:Stokes}
    St = t_{s} \Omega = \frac{\pi}{2} \frac{a \rho_{d}}{\Sigma_{g}}.
\end{equation} 

We then use the result of the hydrodynamical simulations to interpolate the surface density linearly in terms of Stokes number. 
If the calculated Stokes number is smaller than the smallest value for which there is a hydrodynamical simulation ($St = 2 \times 10^{-3}$) then the dust is assumed to follow the gas surface density distribution. 
The largest grains in the model discs considered typically have $St \sim 0.0037$ at the inner edge, and $St \sim 0.4$ at their outer edge, so the hydrodynamical simulations provide sufficient coverage for the surface density of these particles to be constructed. 
The mass in each dust grain size bin was scaled according to equation \ref{eq:size_dist}. 
The $3$D density profiles are calculated from the surface density as:
\begin{equation}
	\label{eq:3Ddens}
	\rho (r) = \frac{\Sigma (r, \phi) }{\sqrt{2 \pi} H(r)} \mathrm{e}^{\frac{-z^{2}}{2 H(r)^{2}}},
\end{equation}
where $z = r \cos(\theta)$ is vertical height within the disc and $H (r)$ is the pressure scale height as a function of position in the disc, calculated as:
\begin{equation}
	\label{eqn:H}
    H (r) = h_{0} \bigg( \frac{r}{r_{0}} \bigg)^{0.25} r,
\end{equation}
where $h_{0}$ is the reference aspect ratio taken at $r_{0}$, the planet location. 

\subsubsection{RADMC-$3$D Parameters}
The radiative transfer simulations and image generation were carried out using RADMC-$3$D. 
All images were calculated at a wavelength of $850 \, \mathrm{\mu m}$, equivalently a frequency of $353 \, \mathrm{GHz}$ which corresponds to ALMA band 7. 
The radiative transfer simulations used $2 \times 10^{7}$ photons, and $1 \times 10^{7}$ photons were used for the image generation. 
We find that this number of photons is sufficiently high to show little noise in the resulting images. 

\subsection{Gap Analysis} \label{sssec:gapanalysis}
There have been several different methods for characterising gap properties proposed in the literature (\citet{DeJuanOvelar2013} and \citet{Akiyama2016}). 
In this work we modify the definition for depth described by \citet{Rosotti2016}, as described here. 
The data from the simulated images was averaged azimuthally, to give a radial surface brightness profile, $S_{\nu}(r)$. 
All discs were assumed to be at a distance of $140 \, \mathrm{pc}$ and face on. 

If the surface brightness profile shows an obvious gap feature then there is no need for further analysis. 
In some cases there is a less distinctive feature visible and therefore a more robust definition of whether a gap exists is needed. 
In order to determine the detectability of a gap a linear fit in log-log space was applied to a region of the surface brightness profile near the feature to calculate a background surface brightness profile, $S_{\nu,b}(r)$. 
The depth of a potential gap is defined to be:
\begin{equation}
    \label{eqn:depth}
    Depth = \bigg{|} \frac{S_{\nu}(r_\mathrm{gap}) - S_{\nu, b}(r_\mathrm{gap})}{S_{\nu, b}(r_\mathrm{gap})} \bigg{|}
\end{equation}
where $r_\mathrm{gap}$ is the location of the gap (where the difference between the real surface brightness profile and the background profile was greatest). 

\begin{figure}
	\centering
    \includegraphics[width = 0.4\textwidth]{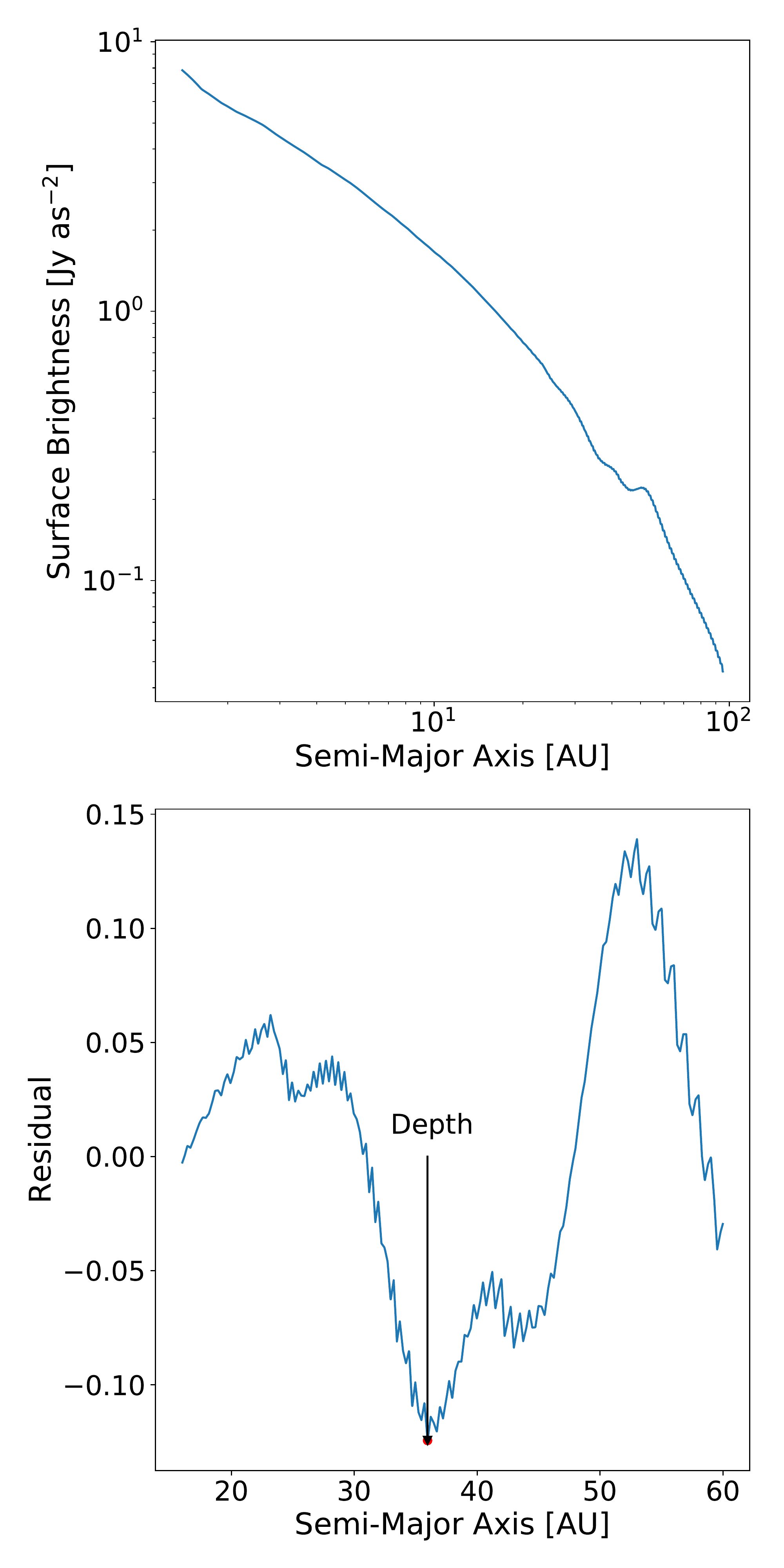}
    \caption{The top plot shows the azimuthally averaged surface brightness profile extracted from the simulation of a $20 \, \mathrm{M_{\earth}}$ mass planet at a semi-major axis of $36.2 \, \mathrm{AU}$ in the disc around a $1 \, \mathrm{M_{\sun}}$ star. 
    There is a feature visible at approximately $40 \, \mathrm{AU}$ but it is not definitively a gap. 
    The bottom plot shows the results of the analysis carried out on the normalised surface brightness profile, with the depth of the gap marked by a black arrow. 
    This corresponds to a depth of $\sim 13 \% $ and so the gap is defined as detectable. }
	\label{fig:20_066_Gap_properties}
\end{figure}

We define gaps as detectable if $ Depth \geq 0.1$, i.e. if the decrease in surface brightness is greater than $10\%$. 
This method is illustrated in figure \ref{fig:20_066_Gap_properties}, for the case of a system containing a $20 \, \mathrm{M_{\earth}}$ mass planet in the disc around a $1 \, \mathrm{M_{\sun}}$ mass star at semi-major axis of $36.2 \, \mathrm{AU}$. 
The depth of the gap is marked and the absolute change in surface brightness was found to be greater than the limit adopted, therefore the feature was defined as a gap. 

\subsection{Simulated observations}\label{ssec:CASA}
We use images produced by the radiative transfer code to generate simulated observations of the system, using the Common Astronomy Software Application\footnote{\url{http://casa.nrao.edu/index.shtml}} (CASA) v5.1.2-4. 
The \verb+simobserve+ task was used to simulate the observed visibilities, from which the simulated observations were produced using the \verb+simanalyse+ task. 
The full $12 \, \mathrm{m}$ array was used in configuration $24$, which gave a resolution at $850 \, \mathrm{\mu m}$ of approximately $0.025 \arcsec$. 
We assume an integration time of $6 \, \mathrm{hours}$, and use the full bandwidth of $7.5 \, \mathrm{GHz}$. 
Noise was introduced using the \verb+tsym-atm+ parameter, with the value for the precipitable water vapour, $0.913 \, \mathrm{mm}$, representative of typical observing conditions.\footnote{\url{https://almascience.eso.org/proposing/sensitivity-calculator}}
As before, all discs are assumed to be at a distance of $140 \, \mathrm{pc}$. 
A gap was defined as detectable in the simulated images in the same way as for radiative transfer images, described in section \ref{sssec:gapanalysis}. 

\section{Results}\label{sec:results} 
\subsection{A Single Representative System}
\begin{figure*}
	\centering
    \includegraphics[width=0.85\textwidth]{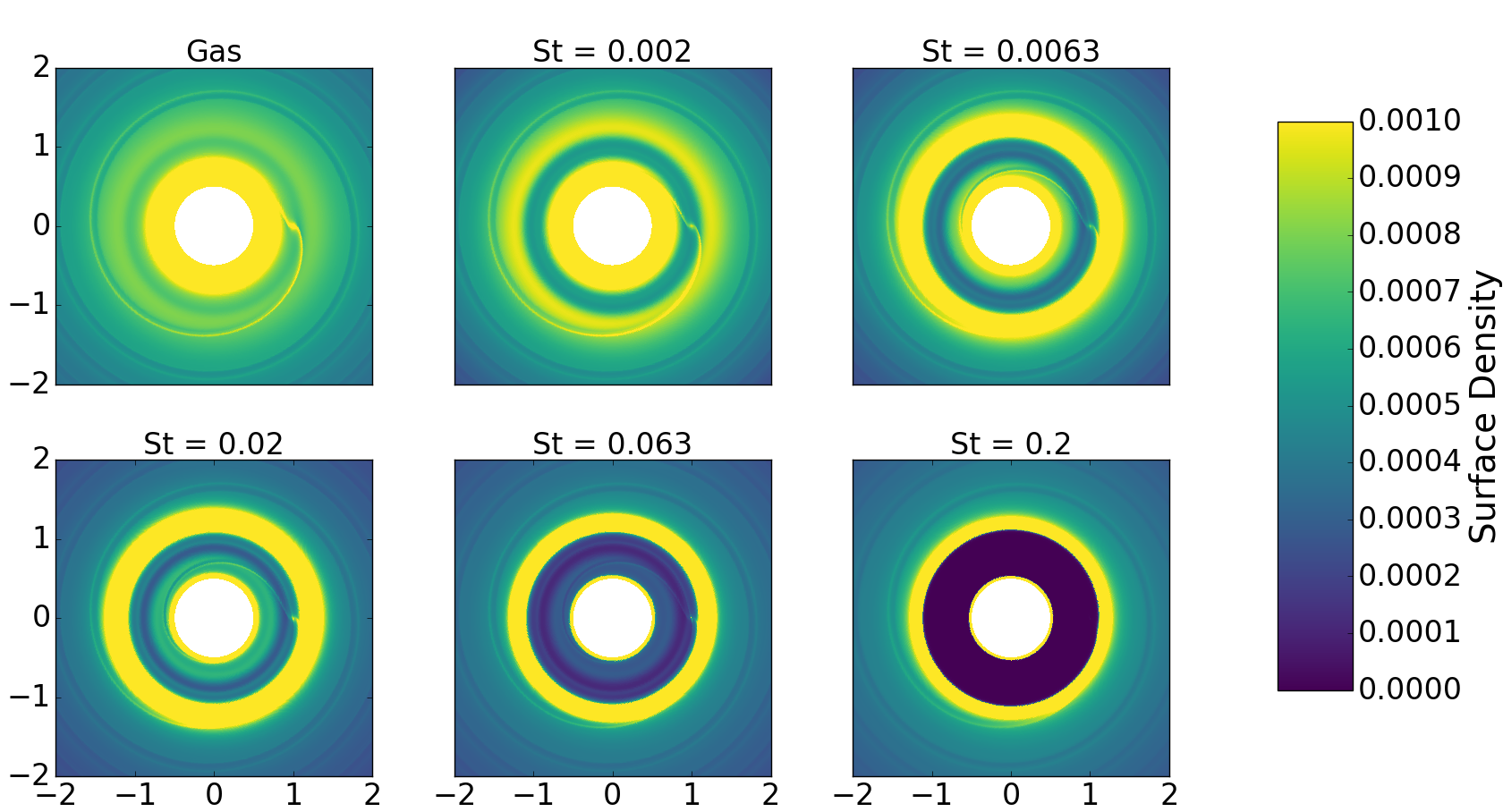}
    \caption{Surface density plots of the FARGO simulation data for an example disc, in this case for a $20 \, \mathrm{M_{\earth}}$ mass planet for a $1 \, \mathrm{M_{\sun}}$ mass star, and an aspect ratio at unit radius of $0.05$.
    The surface density profiles for the gas and the five different Stokes numbers are shown. 
    The planet opens a shallow gap in the gas, which is deeper and wider for increasing Stokes number. 
    Spiral features are also visible, and are more pronounced in the gas and dust species with low Stokes numbers. 
    In the dust species with the largest Stokes number the planet opens a hole, which extends from the location of the planet to the inner edge of the disc (at half unit radius). }
    \label{fig:FARGO_Sig}
\end{figure*}

The methodology described in section \ref{sec:methods} is illustrated here for the case of a $20 \, \mathrm{M_{\earth}}$ mass planet around a $ 1 \, \mathrm{M_{\sun}}$ star at the location where the aspect ratio is $0.05$, which corresponds to a semi-major axis of $19.8 \, \mathrm{AU}$. 
The FARGO surface density data for this simulation is shown in figure \ref{fig:FARGO_Sig}. 
From this figure it can be seen that the dust surface density profiles are largely azimuthally symmetric, with the exception of a thin spiral feature that is more pronounced in the gas and dust species with low Stokes numbers. 
Previous work by \cite{2015MNRAS.451.1147J} has suggested that, even for extremely massive planets, spiral features may be challenging to observe in continuum images as they are narrow and have low contrast. 
As a consequence, we use azimuthally averaged profiles of both surface density and image surface brightness in the remainder of this work without further detailed consideration of any asymmetric features. 

\begin{figure}
	\centering
    \includegraphics[width = 0.5\textwidth]{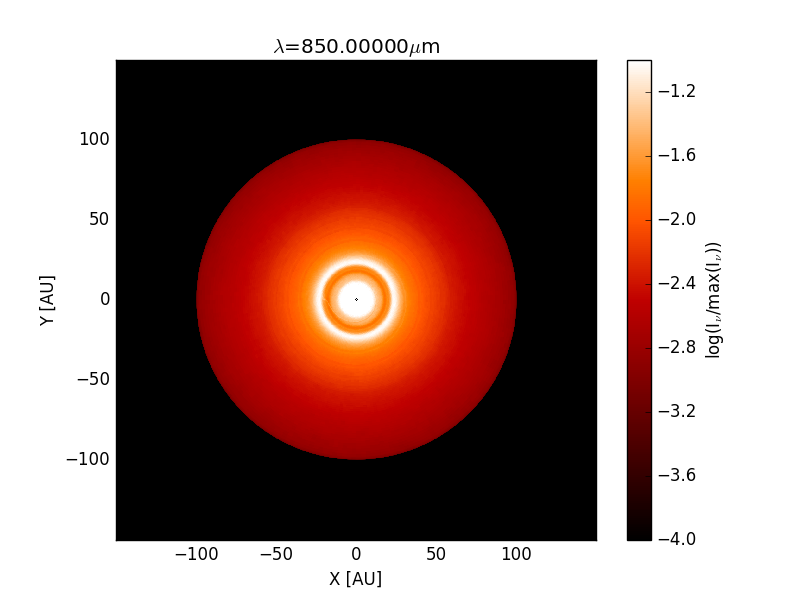}
    \caption{The radiative transfer image of a model disc around a $1 \, \mathrm{M_{\sun}}$ star containing a $20 M_{\earth}$ mass planet at an aspect ratio of $0.05$, corresponding to a semi-major axis of $19.8 \, \mathrm{AU}$. }
	\label{fig:Image}
\end{figure}

\begin{figure*}
    \centering
    \begin{minipage}{0.45\textwidth}
        \centering
        \includegraphics[width=1.\textwidth]{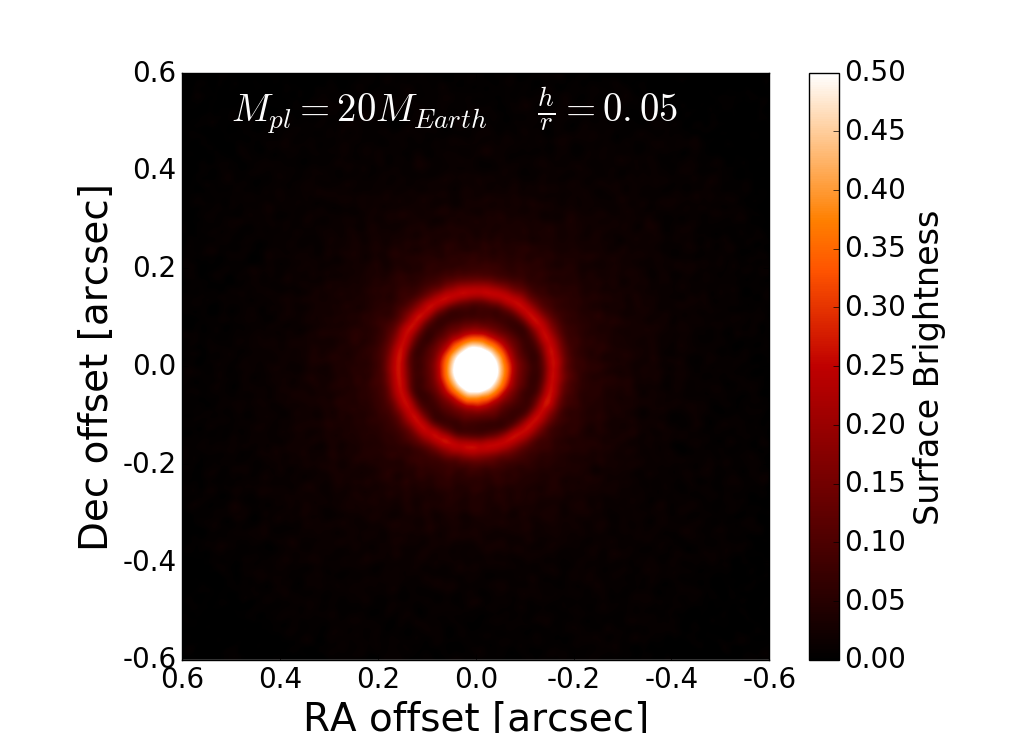}
        \caption{Simulated observation in ALMA band $7$, at $850 \, \mathrm{\mu m}$, of a model disc around a $1 \, \mathrm{M_{\sun}}$ star containing a $20 M_{\earth}$ mass planet at an aspect ratio of $0.05$, and therefore a semi-major axis of $19.8 \, \mathrm{AU}$. 
    The disc is assumed to be face on at a distance of $140 \, \mathrm{pc}$. 
    It should be noted that the colour scale in this image is not the same as that used in figure \ref{fig:Image}}
		\label{fig:Sim_Obs}
    \end{minipage}\hfill
    \begin{minipage}{0.45\textwidth}
        \centering
        \includegraphics[width=0.83\textwidth]{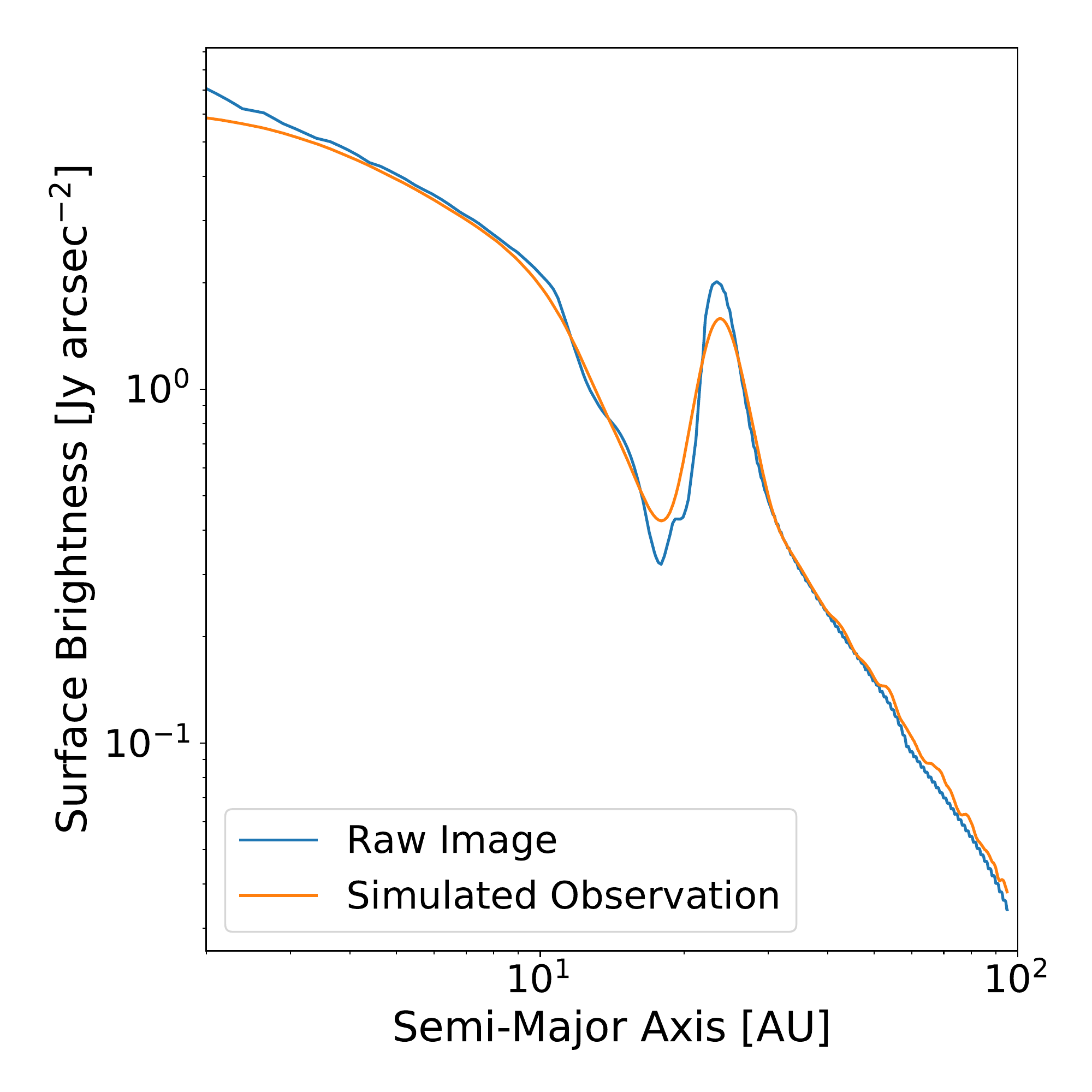}
        \caption{Azimuthally averaged surface brightness profiles produced from the radiative transfer image and simulated observation of a model disc around a $1 \, \mathrm{M_{\sun}}$ star containing a $20 M_{\earth}$ mass planet at an aspect ratio of $0.05$, and therefore a semi-major axis of $19.8 \, \mathrm{AU}$.}
    	\label{fig:casa_Sb}
    \end{minipage}
\end{figure*}

The image produced by RADMC-$3$D is shown in figure \ref{fig:Image}, and the corresponding simulated observation is shown in figure \ref{fig:Sim_Obs}. 
The gap created by the planet can clearly be seen, as can the bright ring produced outside the location of the planet. 

The corresponding azimuthally averaged surface brightness profiles extracted from these images are shown in figure \ref{fig:casa_Sb}. 
Both show a clear decrease in surface brightness at the location of the planet below the background level. We can thus conclude that this planet is gap opening without the need for further analysis. The thermal noise can be more clearly seen in the outer portion of the disc, where the surface brightness is low. 

\subsection{Summary of $1 \, \mathrm{M_{\sun}}$ Mass Star Case}\label{ssec:1Ms}
For a disc around a $1 \, \mathrm{M_{\sun}}$ mass star (for which $R_{d} = 100 \, \mathrm{AU}$), images at $850 \, \mathrm{\mu m}$ were simulated for discs containing planets with masses of $2.5$, $4$, $8$, $12$, $20$, $60$ and $120 \, \mathrm{M_{\earth}}$ at six orbital radii between $2.5$ and $75.7 \, \mathrm{AU}$. 
Radiative transfer images and simulated observations were generated and analysed as described in sections \ref{ssec:RT}, \ref{sssec:gapanalysis} and \ref{ssec:CASA}. 

\begin{figure*}
    \centering
    \begin{minipage}{0.45\textwidth}
        \centering
        \includegraphics[width=0.85\textwidth]{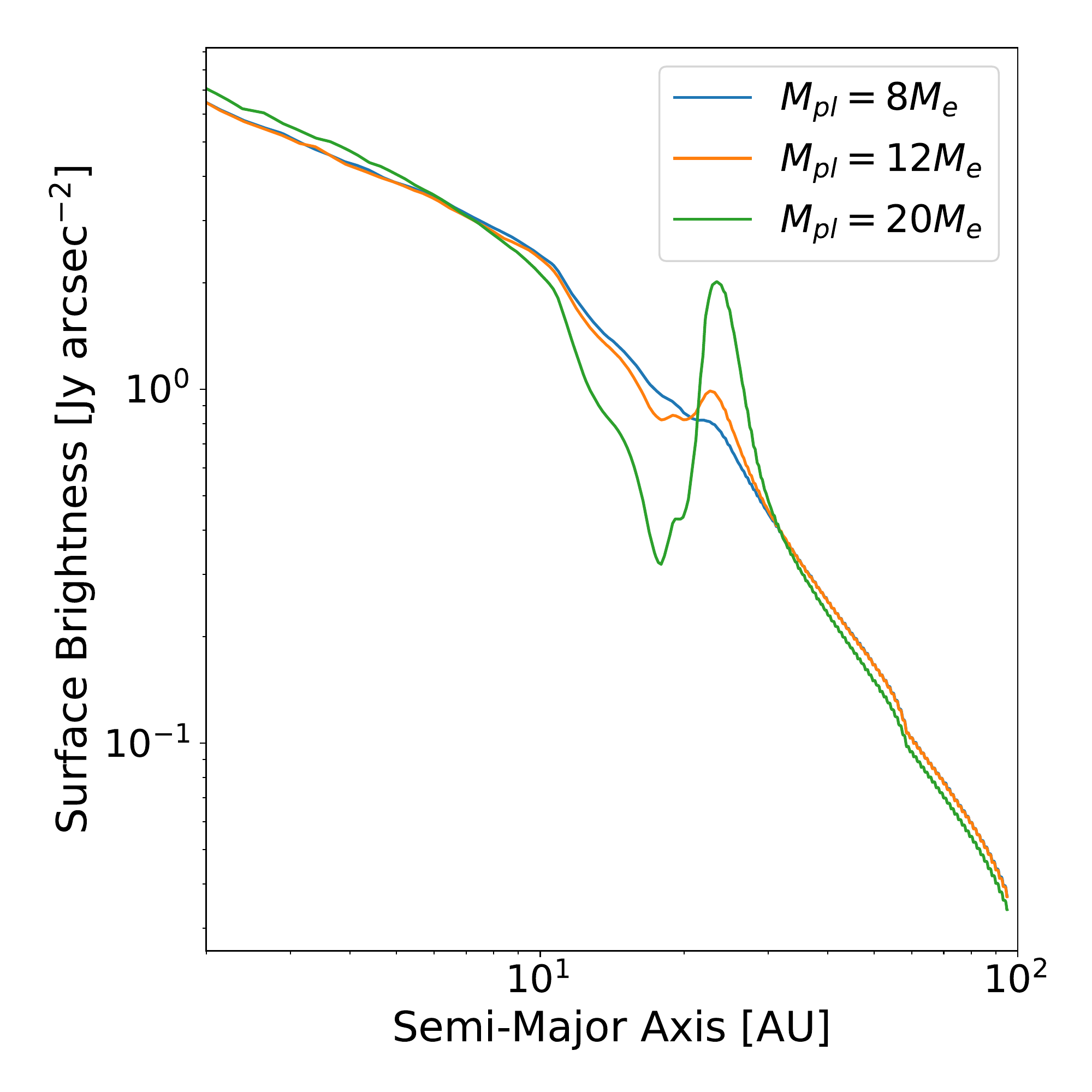}
        \caption{Azimuthally averaged surface brightness profiles showing the effect of varying the planet mass for a fixed semi-major axis. 
    These profiles were generated from the images produced by RADMC-$3$D of three model discs around a $1 \, \mathrm{M_{\sun}}$ star containing three different mass planets ($8$, $12$ and $20 \, \mathrm{M_{\earth}}$) at an aspect ratio of $0.05$, and therefore a semi-major axis of $19.8 \, \mathrm{AU}$. More massive planets create more notable feature in the surface brightness profile.}
		\label{fig:hr05_Sb}
    \end{minipage}\hfill
    \begin{minipage}{0.45\textwidth}
        \centering
        \includegraphics[width=0.85\textwidth]{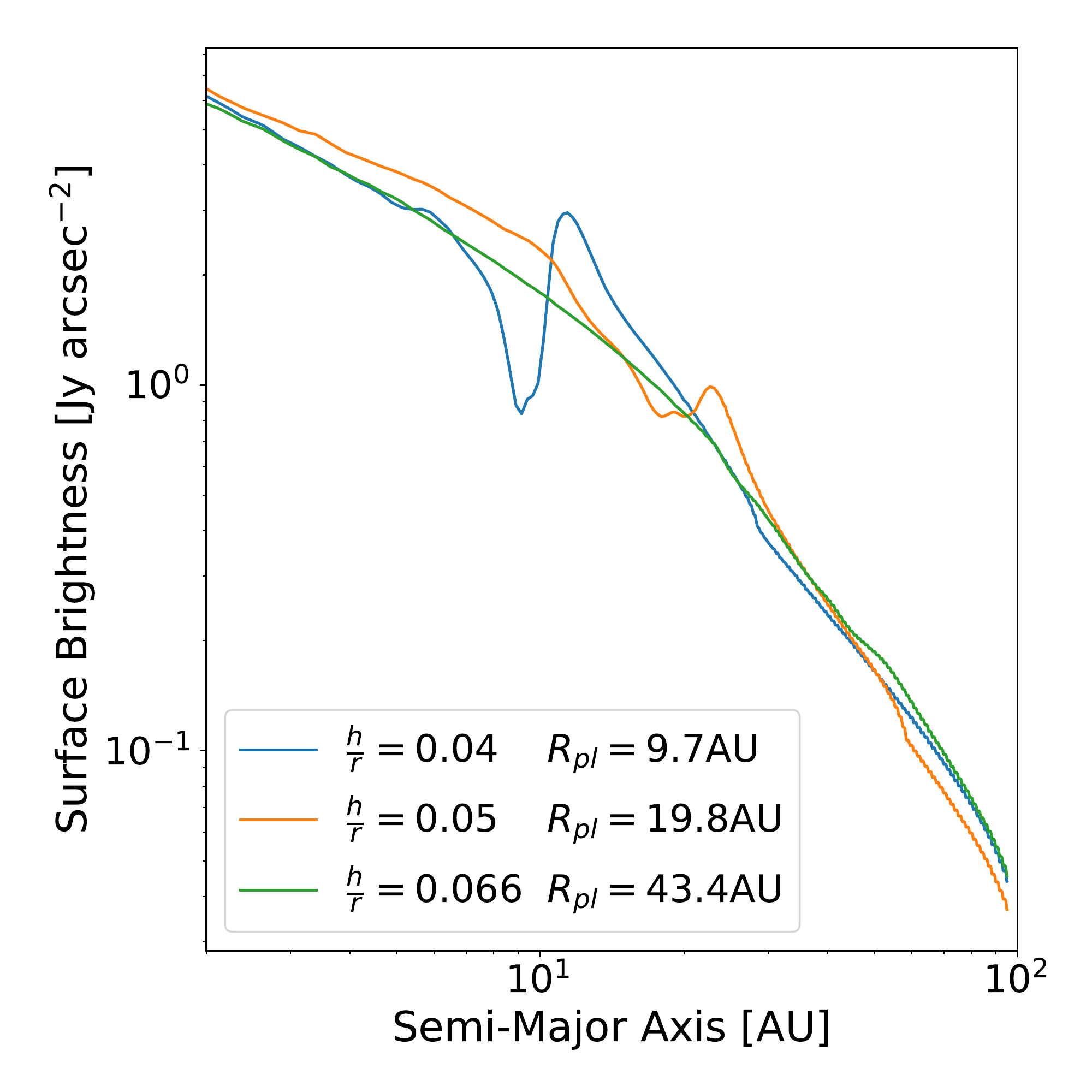}
        \caption{ Illustrative plot showing the effect of varying the planet semi-major axis for a fixed planet mass. The surface brightness profiles shown are for the case of a $12 \, \mathrm{M_{\earth}}$ mass planet around a $1 \, \mathrm{M_{\sun}}$ star, but at three different locations. The perturbation induced by the planet becomes stronger at smaller radii, as a result of the lower disc aspect ratio. }
		\label{fig:12Me_Sb}
    \end{minipage}
\end{figure*}

\paragraph*{Radiative Transfer Images}
For a given semi-major axis more massive planets create more obvious features within the disc. 
This is illustrated in figure \ref{fig:hr05_Sb}, which shows the azimuthally averaged surface brightness profiles for three discs containing planets of different masses at $19.8 \, \mathrm{AU}$. 
The most massive planet, with a mass of $20 \, \mathrm{M_{\earth}}$, produces a large decrease in surface brightness, as well as a bright ring outside the location of the planet. 
These same features can be seen but are much less pronounced for the smaller planet, with a mass of $12 \, \mathrm{M_{\earth}}$, and no obvious feature can be seen at all for the least massive, $8 \, \mathrm{M_{\earth}}$ mass planet. 

For a planet of given mass, the features produced are more prominent for planets located at smaller semi-major axes, as illustrated in figure \ref{fig:12Me_Sb}. 
This is due to the lower aspect ratio at smaller semi-major axes, for which a lower planet mass is required to open a gap. 
This plot shows the azimuthally averaged surface brightness profiles produced from the simulations of three discs containing $12 \, \mathrm{M_{\earth}}$ mass planets at different semi-major axes. 
A deep gap and prominent bright ring is produced by the planet at $9.7 \, \mathrm{AU}$, while a smaller gap is produced by the planet at $19.8 \, \mathrm{AU}$. 
The planet furthest out, at $36.2 \, \mathrm{AU}$, produces no visible feature at all. 

\begin{figure} 
	\centering
    \includegraphics[width = 0.4\textwidth]{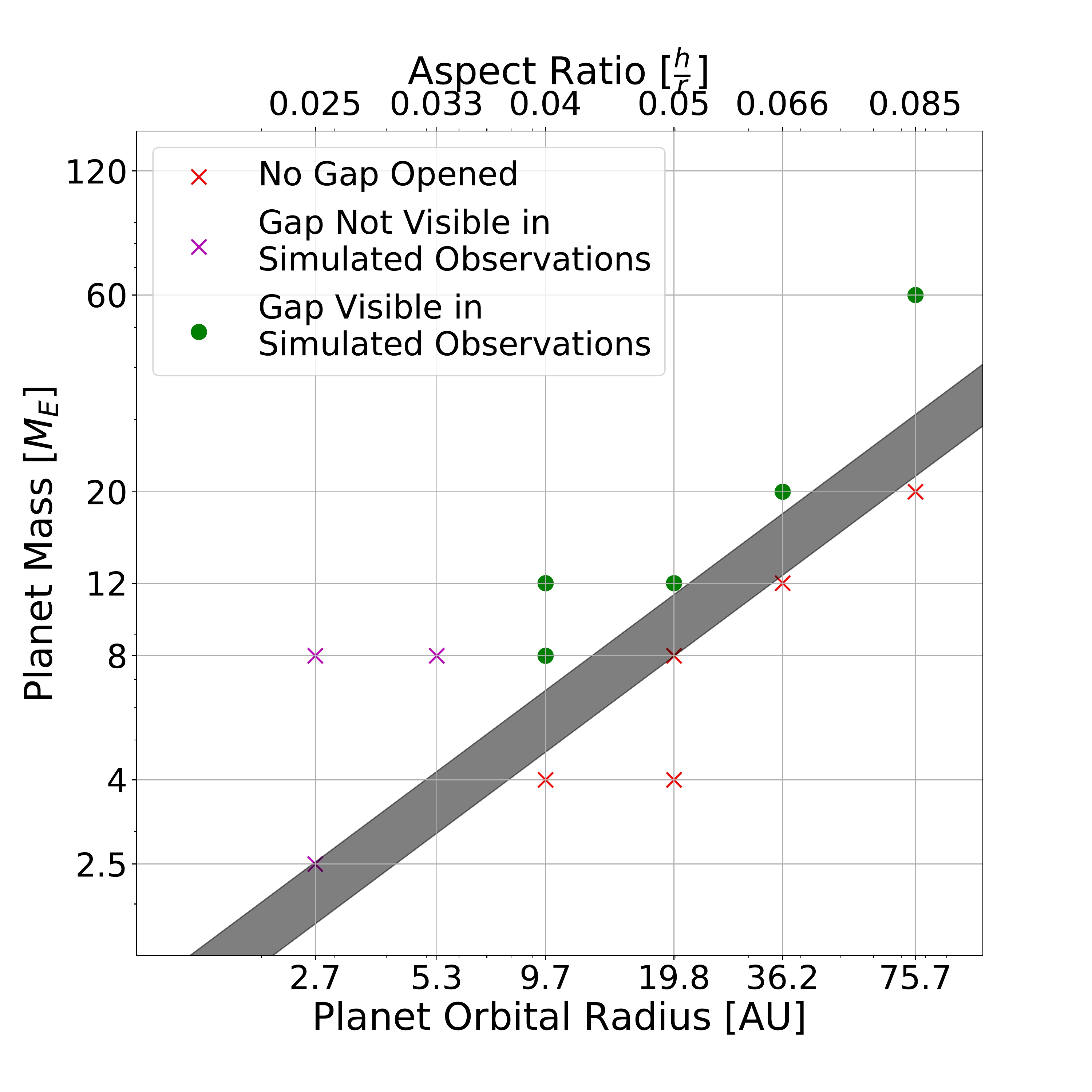}
    \caption{A summary of the detectability of gaps created by planets within the disc around a $1 \, \mathrm{M_{\sun}}$ star. 
    A red cross indicates that no detectable gap was produced. 
    A purple cross indicates that a detectable gap was visible in the radiative transfer image, but not in the simulated observation. 
    A green circle indicates a detectable gap that was visible in both the radiative transfer image and the simulated observation. 
    The grey shaded region indicates the $M_{min} \propto r^{0.75}$ fit supported by these results, which is in line with the theoretical arguments we present in section \ref{ssec:expected_scalings}. }
	\label{fig:Summary_10}
\end{figure}

\paragraph*{Simulated Observations}
For each case in which a detectable gap was opened a simulated ALMA band $7$ observation was generated, as described in section \ref{ssec:CASA}. 
From these we recover the same trends discussed in the images. 
\paragraph*{Results}
The results for all simulations performed for the $1 \, \mathrm{M_{\sun}}$ case are summarised in figure \ref{fig:Summary_10}, which distinguishes between three different results: $1)$ no gap is present in the radiative transfer image, $2)$ a gap is detected in the radiative transfer image but is not visible in the simulated observation, $3)$ a gap is detectable in both the radiative transfer image and the simulated observation. 

This summary figure shows that gaps are opened for small semi-major axes and/or large planet mass. 
We find that the noise introduced by simulating an ALMA observation has little effect on the detectability of gaps. 
A more important effect is the finite resolution of the simulated observations. 
For planets with semi-major axes of $5.3 \, \mathrm{AU}$ or less, the gap that was defined as detectable in the radiative transfer image is no longer visible in the data extracted from the simulated observation. 
The ALMA configuration we used gave an angular resolution of $0.025 \arcsec$, which at a distance of $140 \, \mathrm{pc}$ is approximately $ 5 \, \mathrm{AU}$, explaining this result. 

\subsection{Changing Stellar Mass}
\begin{figure} 	
	\centering
    \includegraphics[width = 0.4\textwidth]{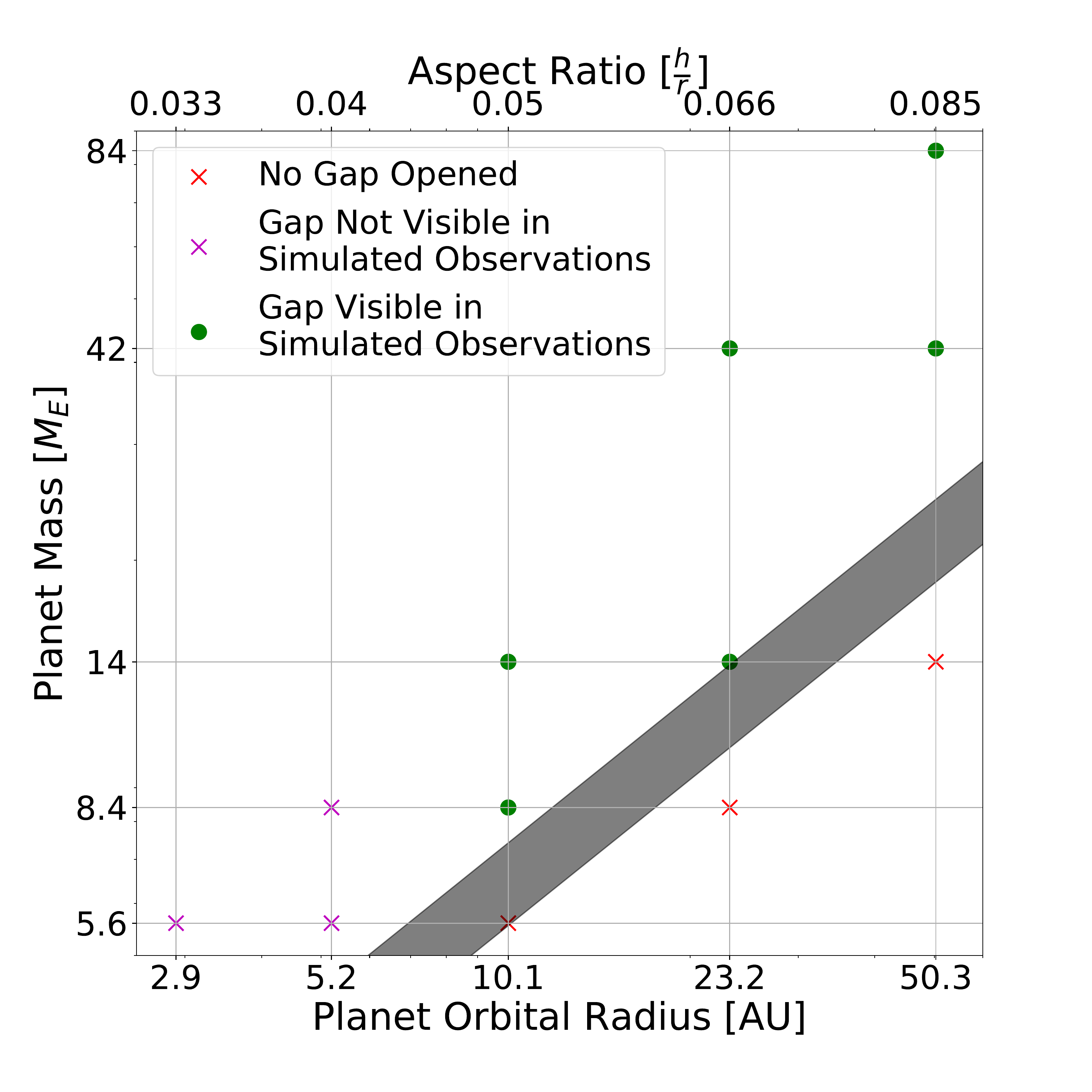}
    \caption{A summary of the detectability of features created by planets within the disc around a $0.7 \, \mathrm{M_{\sun}}$ mass star. See figure \ref{fig:Summary_10} for an explanation of colours and symbols.}
	\label{fig:Summary_07}
\end{figure}

\begin{figure} 	
	\centering
    \includegraphics[width = 0.4\textwidth]{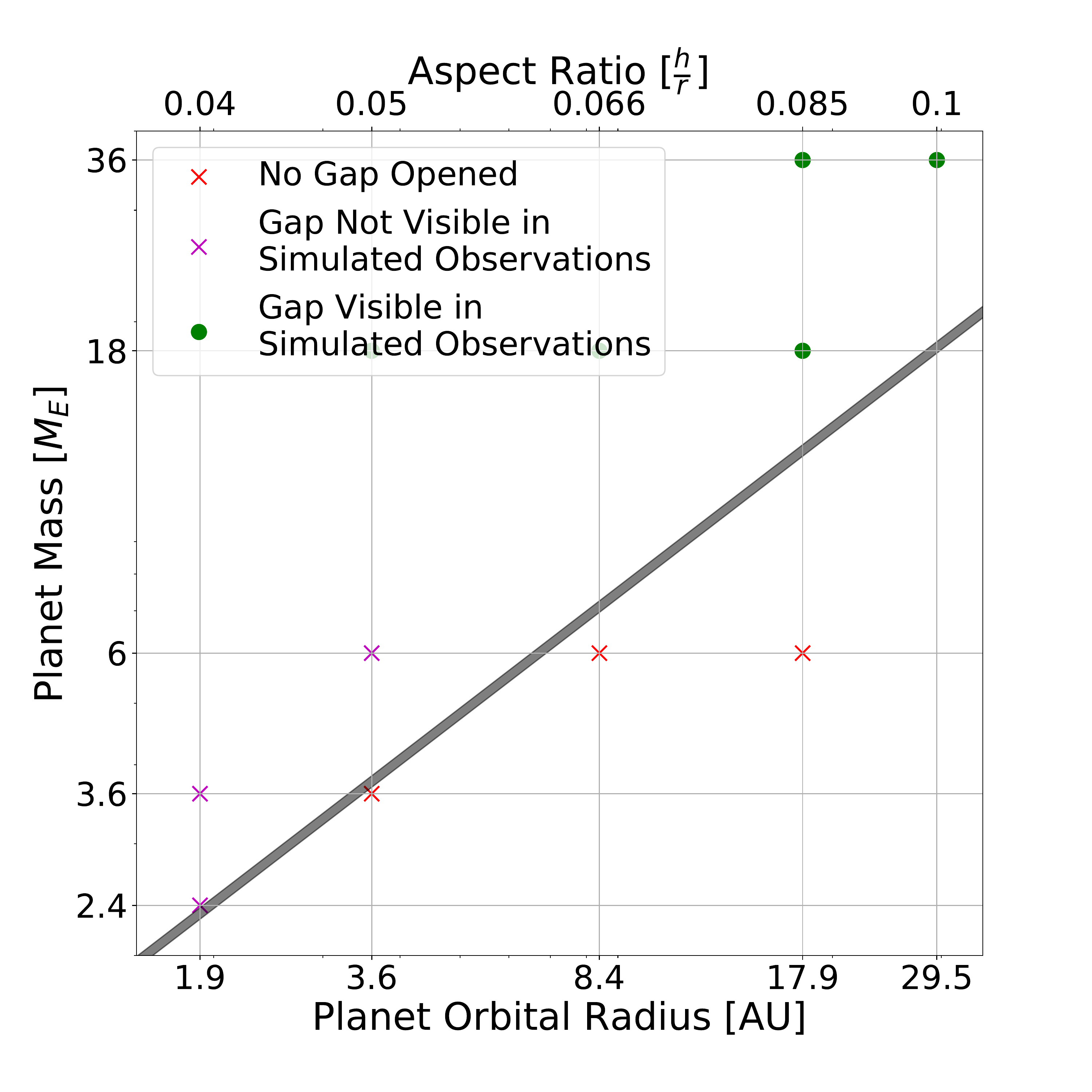}
    \caption{A summary of the detectability of features created by planets within the disc around a $0.3 \, \mathrm{M_{\sun}}$ mass star. See figure \ref{fig:Summary_10} for an explanation of colours and symbols.}
	\label{fig:Summary_03}
\end{figure}

For a disc around a $ 0.7\, \mathrm{M_{\sun}}$ mass star ($R_{d} = 81 \, \mathrm{AU}$), observations at $850 \, \mathrm{\mu m}$ were simulated for discs containing planets at six orbital radii between $2.9$ and $50.3 \, \mathrm{AU}$. 
The images generated by the radiative transfer code and the simulated observations were analysed as discussed in sections \ref{sssec:gapanalysis} and \ref{ssec:CASA} with the results shown in figure \ref{fig:Summary_07}. \\
For a disc around a $ 0.3\, \mathrm{M_{\sun}}$ mass star ($R_{d} = 48 \, \mathrm{AU}$), observations were simulated for discs containing planets at orbital radii between $1.9$ and $29.5 \, \mathrm{AU}$. 
The images generated were analysed as discussed in sections \ref{sssec:gapanalysis} and \ref{ssec:CASA} with the results are shown in figure \ref{fig:Summary_03}.

In both of these systems, as for the $1 \, \mathrm{M_{\sun}}$ case, the gaps at small semi-major axes visible in the radiative transfer images are not visible in the simulated observations. 
Inspection of these simulated images suggests qualitatively that it is harder to open gaps in discs around lower mass stars. 
For example, the minimum planet mass required to produce a visible gap at $\sim 10 \, \mathrm{AU}$ in the disc around a $1 \, \mathrm{M_{\sun}}$ mass star is approximately $5 - 6 \, \mathrm{M_{\earth}}$, but in the disc around a $0.3 \, \mathrm{M_{\sun}}$ mass star it is approximately $8 \, \mathrm{M_{\earth}}$. 
We discuss this result in section \ref{ssec:comparison}. 

\section{Dependence on the stellar mass}
\label{sec:discussion}

We now investigate the variation in the MGOPM with, most importantly, the stellar mass, and also with semi-major axis within the disc. To do this we assume that the dependence on these two parameters is a power law and we fit the results presented in sections \ref{sec:results} to derive the exponents. We also present analytic scaling arguments and compare these to our results. We then discuss the robustness of our results, considering especially the effect of the luminosity spread, and finally discuss the observational implications.

\subsection{Results of the numerical simulations} \label{ssec:comparison}
The exponent of the power law relating MGOPM and semi-major axis is compatible with a value of $0.75$, and therefore we use this value in the following analysis. We will show in section \ref{ssec:expected_scalings} that this value is in agreement with theoretical arguments.

Figures \ref{fig:Summary_10}, \ref{fig:Summary_07} and \ref{fig:Summary_03} show the fits for the three different stellar masses considered by this work. We also show as a gray shaded region the allowed range of normalisation constants that are in agreement with our results. In all cases this power law provides a plausible fit. 
The range of allowed normalisation constants is small for the $0.3 \, \mathrm{M_{\sun}}$ mass star, and larger for the $0.7$ and $1 \, \mathrm{M_{\sun}}$ mass stars. 
In general the relation between MGOPM and semi major axis for a given stellar mass can be expressed as:
\begin{equation}
\begin{aligned}
& & & & 1.41 \leq \, &A_{0.3 \mathrm{M_{\sun}}} \leq 1.47 \\
\frac{M_{pl,m}}{\mathrm{M_{\earth}}} &= A_{M_{*}} \bigg( \frac{r_{pl}}{\mathrm{AU}} \bigg)^{0.75} & &\mathrm{where} & 0.98 \leq \, &A_{0.7 \mathrm{M_{\sun}}} \leq 1.31 \\
& & & & 0.85 \leq \, &A_{1.0 \mathrm{M_{\sun}}} \leq 1.2
\label{eqn:minplanetmass}
\end{aligned}
\end{equation}

Addressing the main motivation behind this paper, from these fits it can be seen that $A_{M_{*}}$ \textit{increases} with decreasing stellar mass. This means that the fact that these discs are geometrically thicker is the dominant effect. We will show this more formally in the next section \ref{ssec:expected_scalings}. The scaling of the MGOPM with stellar mass is consistent with:
\begin{equation}
\begin{aligned}
A_{M_{*}} &\propto M_{*}^{\alpha} & &\mathrm{where} & -0.45 \leq \, &\alpha \leq -0.14
\end{aligned}
\end{equation}

In summary, the MGOPM is larger at greater semi-major axes for fixed stellar mass, and at fixed semi-major axes is larger for lower mass stars. 
Using a representative value of $A_{M_{*}} \approx 1 \, M_{*}^{-0.33}$, the MGOPM can be expressed as a function of stellar mass and planet semi-major axis as:
\begin{equation}
\frac{M_{pl,m}}{\mathrm{M_{\earth}}} \approx 1 \,\left( \frac{M_{*}}{\mathrm{M_{\sun}}} \right)^{-0.33} \left( \frac{r_{pl}}{\mathrm{AU}} \right)^{0.75}
\label{eqn:summary_Mmin}
\end{equation}

\begin{figure} 	
	\centering
    \includegraphics[width = 0.4\textwidth]{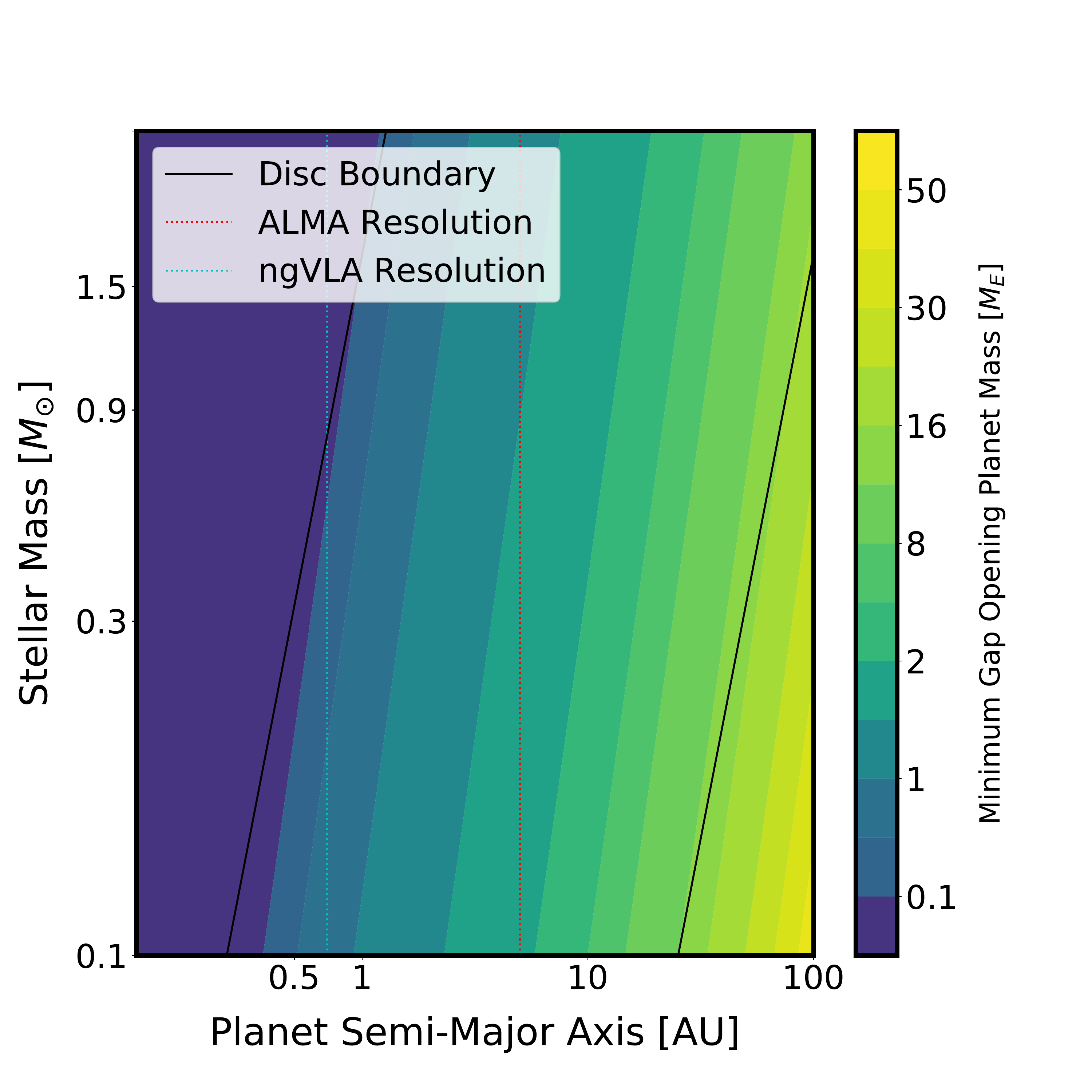}
    \caption{A summary of the minimum gap opening planet mass (MGOPM) as a function of planet semi-major axis and stellar mass. The black line is the average size of a disc for the given stellar mass.
    The estimated resolution of the ALMA configuration used ($\sim 5 \, \mathrm{AU}$) is shown as a red dashed line. 
    The estimated resolution of $3 \, \mathrm{mm}$ observations using the ngVLA ($\sim 0.7 \, \mathrm{AU}$) is shown as a blue dashed line. It can be seen that the MGOPM increases with semi major axis and decreases with stellar mass.
}
	\label{fig:summary}
\end{figure}

The relation given in equation \ref{eqn:summary_Mmin} is illustrated in figure \ref{fig:summary}. The plot also shows the resolution of the simulated observations, $\sim 5 \, \mathrm{AU}$. In addition, while this paper focuses on ALMA, for reference we plot also the resolution that may be achieved using the ngVLA\footnote{https://science.nrao.edu/futures/ngvla} at $3 \, \mathrm{mm}$ ($5$ milliarcsec, which at a distance of $140 \, \mathrm{pc}$ corresponds to a distance of $0.7 \, \mathrm{AU}$). See \citet{Ricci2018} for a dedicated study of ngVLA capabilities in detecting planet-formed gaps.

It can be seen how, at least on average, around a lower mass star a smaller part of the disc can be resolved by ALMA (e.g., for a 0.3 $M_\odot$ star the average outer radius is 30 au, i.e. the dynamical range in radius is a factor of 6). In this region, the MGOPM is approximately that of Neptune (15-20 $M_\oplus$). Around solar mass stars, the negative scaling with stellar mass means that the sensitivity improves, and the detection limit becomes $\sim 5 M_\oplus$.


\subsection{Analytical expectations for the MGOPM scaling relations} \label{ssec:expected_scalings}

While the MGOPM we consider in this paper is quantitatively different from the conventional gap opening criterion in the gas (\citealt{Crida2006}; see e.g. \citealt{2014ApJ...782...88F,2015MNRAS.448..994K} for recent developments), the formation of a dust gap still requires a perturbation in the gas. For this reason, although the amplitude of the perturbation is different, we can assume that the two criteria should scale in the same way with the disc parameters. A similar assumption was  also made by \citet{2017MNRAS.469.1932D} following the results of \citet{Rosotti2016}, see for example the blue area in their Figure 2. In this section we validate this assumption, by showing that it accounts for the scalings found in the simulations we have run. For what concerns the normalization, instead, we must rely on numerical simulations.

Proto-planetary discs are characterised by relatively low values of the viscosity and high values of the aspect ratio. Therefore, it is safe to assume that the most stringent criterion for gap opening is the pressure criterion, rather than the viscous one \footnote{This is different in the regime of low viscosity $\alpha \lesssim 10^{-4}$, where the \citet{Crida2006} criterion is no longer applicable \citep[e.g.,][]{2013ApJ...769...41D}.}. For this reason in the rest of the section we will consider only the former (see also discussion in \citealt{Rosotti2016}). Following standard arguments \citep{Lin1993}, an estimate of the expected scaling of the MGOPM $(M_{pl,m})$ with semi-major axis and stellar mass can be obtained by equating the Hill radius, 
\begin{equation}
r_{H}= r \Big( \frac{M_{pl,m}}{3M_{*}} \Big)^{1/3},
\label{eqn:hill_r}
\end{equation}
of the planet with the pressure scale height $(H)$. This gives the scaling condition that:
\begin{equation}
\frac{M_{pl,m}}{M_{*}} \propto \bigg( \frac{H}{r} \bigg)^{3}.
\label{eqn:Mmin}
\end{equation}

For fixed stellar mass, recalling the expression for the aspect ratio given by equation \ref{eqn:H}, this gives:
\begin{equation}
\frac{M_{pl,m}}{M_{\earth}} \propto \bigg( \frac{r}{\mathrm{AU}} \bigg)^{0.75}.
\label{eqn:Mmin_r}
\end{equation}

And for a given position in the disc, using equation \ref{eqn:T_M}, the aspect ratio is expected to scale as: 
\begin{equation*}
    \frac{H}{r} \, \propto \, \sqrt{\frac{T(M_{*})}{\mathrm{K}} \bigg( \frac{M_{*}}{\mathrm{M_{\sun}}}\bigg)^{-1}}
        \, \propto \, \bigg( \frac{M_{*}}{\mathrm{M_{\sun}}}\bigg)^{-0.425}
\end{equation*}

So, using equation \ref{eqn:Mmin}, the MGOPM, for fixed position within the disc is expected to scale with stellar mass as:
\begin{equation}
\label{eq:Mmin_Ms}
    \frac{M_{pl,m}}{\mathrm{M_{\oplus}}} \, \propto \, \bigg( \frac{h}{r_{pl}} \bigg)^{3} \bigg( \frac{M_{*}}{\mathrm{M_{\sun}}}\bigg) \propto \bigg( \frac{M_{*}}{\mathrm{M_{\sun}}} \bigg)^{-0.275}
\end{equation}
Combining these equations \ref{eqn:Mmin_r} and \ref{eq:Mmin_Ms} gives an expression for the expected scaling of the MGOPM with both planet semi-major axis and stellar mass:
\begin{equation}
    \frac{M_{pl,m}}{\mathrm{M_{\oplus}}} \, \propto \, \bigg(\frac{r}{\mathrm{AU}}\bigg)^{0.75} \bigg(\frac{M_{*}}{\mathrm{M_{\sun}}} \bigg)^{-0.275}
    \label{eqn:Mmin_comb}
\end{equation}

Comparing equations \ref{eqn:minplanetmass} and \ref{eqn:Mmin_comb}, we conclude that there is excellent agreement between the results of our simulations and the analytical arguments. As mentioned, in our analysis we have assumed that the exponent of the scaling of the MGOPM with radius is $0.75$, because it gives a good description of the results. The range of exponent values describing the scaling with stellar mass are consistent with the value of $-0.275$ predicted in section \ref{ssec:expected_scalings}.



\subsection{Robustness of the results to changes in the stellar mass luminosity relationship} \label{sec:robustness}

We have demonstrated that, for a particular choice of the relationship
between stellar mass and  luminosity as detailed in Table \ref{tab:stellarprop} (based on the
pre-main sequence evolutionary tracks of  \citet{Siess2000}  at an age of
1Myr), the minimum mass of planets that can be detected at a given radius
in the disc is a decreasing function of stellar mass (equation \ref{eqn:Mmin_comb}).
Thus planet detection via structure in submm images is apparently
harder in the case of lower mass stars.
 This contrasts strongly with the situation encountered in the case of other
planet detection methods. For example, in the case of radial velocity
methods, the  detectable planet mass scales linearly with stellar mass,
whereas for rocky planets (i.e., with a roughly constant density) detected by the transit method it scales as $M_\ast^3$.

We now consider if there is any plausible stellar mass luminosity relation
that could result in a positive dependence of minimum detectable planet mass on stellar mass. Let us assume that, for some mass-luminosity relation, the
scaling of temperature with stellar mass is $T \propto M_\ast^a$ (by analogy
with equation (\ref{eqn:T_M})). Proceeding as in the previous section we obtain that
\begin{equation}
M_{pl,m} \propto M_\ast^{(3a-1)/2}.
\label{eq:mpl_mstar}
\end{equation}

implying that a positive dependence of detectable planet mass on stellar mass would correspond to $a > 0.33$. The temperature (at a given radius) mostly depends on the stellar luminosity, but can also depend on the stellar mass. Therefore, we can parametrise this dependence with the form $T \propto L^b M^c$. If we consider a mass luminosity relation of the form $L \propto M^d$, we obtain

\begin{equation}
    a=c+bd
\end{equation}

\begin{figure} 	
	\centering
    \includegraphics[width = 0.45\textwidth]{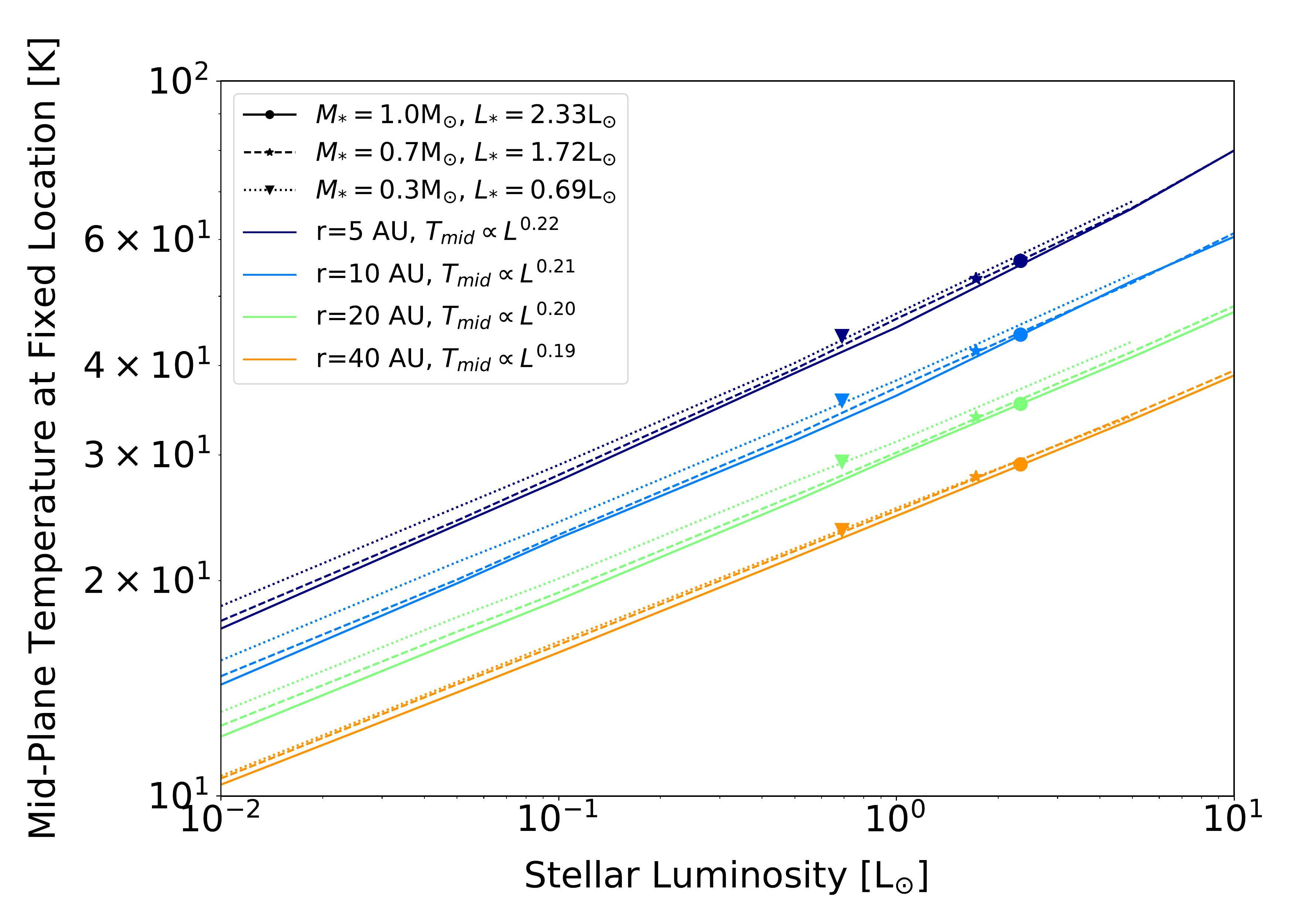}
    \caption{The variation in mid-plane temperature at fixed radial position with stellar luminosity, shown for the three stellar masses and four different locations. From these radiative transfer calculations we deduce a scaling $T_{\rm disc} \propto L_{*}^{0.2}$, slightly flatter than the $L_{*}^{1/4}$ one might naively expect from energy arguments, as a result of the dependence of the Planck mean opacity with temperature (see appendix \ref{sec:append_temp}).}
	\label{fig:midT_fixedr}
\end{figure}

To measure the values of the parameters $b$ and $c$, we have run another grid of radiative transfer simulations of the same type of section \ref{sssec:dischr}, i.e. discs that contain no planets, in which we explored the effect of varying the luminosity of the central star while keeping the other parameters fixed (note the difference from the calculations in section \ref{sssec:stardiscparam}, where we also varied the stellar and disc properties). From these simulations, we find that for a fixed location within the disc, the disc temperature $T_{\rm disc}$ scales with stellar luminosity approximately as
\begin{equation}
\frac{T_{\rm disc}}{\mathrm{K}} \propto \bigg( \frac{L_{*}}{\mathrm{L_{\odot}}} \bigg)^{0.2};
\label{eqn:Tdisk_L}
\end{equation} 
we show the results of our calculations in figure \ref{fig:midT_fixedr}.  Therefore, $b=0.2$. The exponent is slightly flatter than the 1/4 one would naively expect from energy argument; we interpret this as due to the dependence of the Planck mean opacity with temperature (see appendix \ref{sec:append_temp}).

For what concerns the values of $c$, from the radiative transfer grid we deduce a value of $c \sim 0.04$; in an alternative way, we can deduce the value of $c = -0.06$  from the fact that $a=0.15$  (equation (\ref{eqn:T_M})) for the case  $d=1.07$ (appropriate to the Siess et al isochrone at 1 Myr). This means that the explicit dependence on the stellar mass is a small effect, confirming that the temperature is mainly set by the stellar luminosity. 
Neglecting $c$, we obtain that a positive dependence of minimum detectable planet mass on stellar mass  ($a > 0.33$) 
requires $d>1.65$; alternatively, the limiting value is $1.45$ for the case of $c=0.04$. This value is higher than that predicted by the \citet{Siess2000} tracks, in accord with the results of the previous sections. However, for another widely used set of pre-main sequence tracks, the models by \citet{Baraffe2015}, the temperature-luminosity relation is in general steeper; for reference, we find a value of 1.47 at 1 Myr. This value corresponds exactly to the limiting case we identified before. To inspect this case more closely, we extracted the values of the disc temperature from the radiative transfer grid at the stellar luminosities predicted by the \citet{Baraffe2015} tracks. In line with the arguments above, in this case we find a value of $a=0.32$, implying that for the \citet{Baraffe2015} tracks the dependence of MGOPM with stellar mass is essentially flat (see \autoref{eq:mpl_mstar}).

We thus conclude that, depending on the stellar track used, MGOPM could become flat with stellar mass. Nevertheless, we can also conclude that MGOPM \textit{does not improve} towards low stellar masses, in contrast to other planet detection techniques. Therefore, the robust result of this paper is that there is no benefit in terms of planet mass sensitivity when observing discs around lower mass stars.

\subsection{Effect of the luminosity spread} \label{sec:spread}

Up to now we only considered a single luminosity for each stellar mass. In reality, it is well known that in star forming regions stars of the same mass exhibit a wide range of luminosities, a phenomenon colloquially called "luminosity spread", possibly due to the stellar accretion history \citep{2011ApJ...738..140H,2012ApJ...756..118B,2018MNRAS.474.1176J} or an age spread. Since the stellar luminosity affects the disc temperature and therefore MGOPM, we need to quantify how this effect changes the conclusions of this paper.

\begin{figure*} 	
	\centering
    \includegraphics[width = \textwidth]{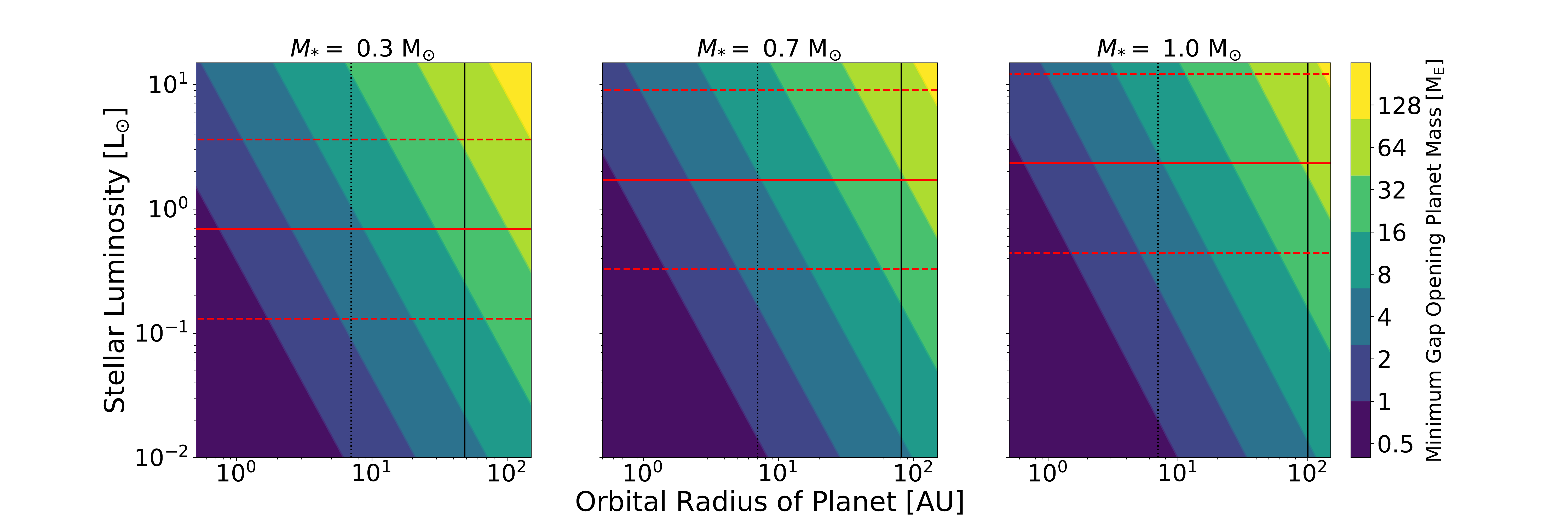}
    \caption{A summary of the minimum gap opening planet mass (MGOPM) as a function of planet semi-major axis and stellar luminosity, for the three values of stellar mass considered. 
    The luminosity values used for each stellar mass are shown by a red solid line; we show with the red dashed lines the typical range of variation (2-$\sigma$, i.e. 95 per cent of the sources) encountered in observations. 
    The average disc size for each mass is shown by a black dashed line.
    The estimated resolution of the ALMA configuration used ($\sim 5 \, \mathrm{AU}$) is shown as a black dotted line. 
    It can be seen that the MGOPM increases with both semi major axis and stellar luminosity.
}
	\label{fig:MGOPM_Ls_apl}
\end{figure*}

To this end, we consider equation \ref{eqn:Mmin} for fixed stellar mass and semi-major axis; we obtain that:
\begin{equation}
\frac{M_{pl,m}}{M_{\earth}} \propto \bigg( \frac{T_{\rm disc}}{\mathrm{K}} \bigg)^{1.5}.
\label{eqn:Mmin_T}
\end{equation}
In the previous section we have shown that $T_{\rm disc} \propto L_\ast^{0.2}$. Combining these gives us an expected power law index for the scaling of the MGOPM with stellar luminosity of $0.3$, and so we expect the scaling with both planet semi-major axis and stellar luminosity to be:
\begin{equation}
    \frac{M_{pl,m}}{\mathrm{M_{\oplus}}} \, \propto \, \bigg(\frac{r}{\mathrm{AU}}\bigg)^{0.75} \bigg(\frac{L_{*}}{\mathrm{L_{\sun}}} \bigg)^{0.3}.
    \label{eqn:Mmin_Ls_apl}
\end{equation}
This predicted MGOPM for three values of the stellar mass are shown in figure \ref{fig:MGOPM_Ls_apl}.

To quantify the importance of the luminosity spread, the last ingredient we need is an estimate of how much the stellar luminosity can vary for a given stellar mass. To quantify this, we have collected the samples presented in the recent X-shooter spectral surveys of Chameleon I \citep{Manara2017} and Lupus \citep{Alcala2017}. We extracted from the two samples the luminosity and stellar mass (note that \citealt{Alcala2017} reports three different values for the stellar mass depending on the model used; here we use only the value derived from the models of \citealt{Baraffe2015} since this is the only one employed by \citealt{Manara2017} and,  as discussed by these authors, there is little difference between the models in deriving the stellar mass) and then fitted them with a power-law using the widely-used package \texttt{linmix} \citep{Kelly2007}. The result of the fit reports a 1-$\sigma$ spread of 0.36 dex. To show this on figure \ref{fig:MGOPM_Ls_apl}, we have indicated with the dashed lines the 2-$\sigma$ spread around the average value.

In itself, the effect of the luminosity spread can be significant: as an illustrative example, a 0.3 $M_\odot$ star with a luminosity that is 2-$\sigma$ below the average has a similar MGOPM (or even smaller) to a 1 $M_\odot$ star with an average luminosity. At the same time, we point out that the luminosity is a quantity that can easily be estimated from optical observations and allows one to correct the estimate of MGOPM for a specific disc.

\subsection{Observational implications}

In this paper we have presented scaling relations of the MGOPM with stellar mass, luminosity and planet orbital radius. These relations, summarised in figure \ref{fig:MGOPM_Ls_apl}, can be readily used when interpreting high-resolution imaging of discs to set a lower limit on the masses of the putative planets responsible for annular structures. 

As a caveat, in this paper we employed a value of the viscous parameter $\alpha = 10^{-3}$. This value was chosen because it is lower than the current upper limits set by direct measurements of the turbulence \citep[e.g.,][]{2018ApJ...856..117F}, but is still in a reasonable range to account for the observed accretion rates onto young stars without invoking other mechanisms for angular momentum transfer, such as disc winds. The precise value of the MGOPM will depend on the value of $\alpha$, but in general we do not expect the trends that we present here to depend on the value of $\alpha$. Another caveat is that we neglected the effect of dust back-reaction on the gas \citep[e.g.,][]{2018ApJ...868...48K,2018ApJ...854..153W,2019arXiv190910526D}, though this is unlikely to change MGOPM since it becomes relevant only in presence of a strong dust accumulation. This requires a planet well above gap-opening mass, at least for the Stokes numbers we simulate here; the situation might change in presence of significantly larger Stokes numbers.

Consideration of how the MGOPM varies as a function of stellar mass is
of considerable interest for the interpretation of the incidence of structure in submm disc images in different stellar mass ranges . From radial velocity surveys it is evident that  
giant planets (loosely defined as being  more massive than Neptune) at distances up to several au are rarer around lower mass stars   \citep{2008PASP..120..531C,2010PASP..122..905J,2013A&A...549A.109B,2014ApJ...791...91C,2015ARA&A..53..409W}\footnote{There is indication \citep{2013A&A...549A.109B,2013ApJ...767...95D,2015ApJ...814..130M}, both from transit and radial velocity surveys, that super-Earths in the innermost au are in fact \textit{more} abundant around low mass stars than around solar.}. While ALMA surveys of disc sub-structure do not overlap in spatial scales with those probed by radial velocity surveys of mature planet populations, it is nevertheless of interest to discover if the incidence of young planets at large radii is also lower in
low mass stars than in higher mass counterparts. 

Our study shows that using canonical relationships between stellar mass and luminosity, the  opening of gaps around low mass stars is harder, or just as difficult. We note that this depends on the distribution of masses and luminosities present in the population; figure \ref{fig:MGOPM_Ls_apl} also shows, based on data from Lupus and Chamaeleon, that the difference in MGOPM due to the luminosity spread can cancel out the effect due to the stellar mass. Thus future studies of the relative incidence of discs with substructure as a function of stellar mass need to be interpreted with care. 
With knowledge of the stellar luminosity on a source by source basis, Figure  \ref{fig:MGOPM_Ls_apl} can be used to assess whether planet formation at young ages and large radii is indeed disfavoured in the vicinity of lower mass stars.  

At the moment, sufficient high resolution imaging data do not exist to make this test; most of the observations of discs around very low mass stars \citep{2014ApJ...791...20R,2016A&A...593A.111T,2016ApJ...819..102V,2018AJ....155...54W} have low spatial resolution. 
Encouragingly, some high-resolution observations are taking place, e.g. \citet{2018A&A...615A..95P} for a 0.1-0.2 $M_\odot$ star. 
The samples of \citet{Andrews2018} and \citet{2019ApJ...882...49L} also contain a few low-mass stars, though at the moment there is no correlation between the sub-structure properties with the stellar properties \citep{2018ApJ...869L..42H}. 
At the moment it is difficult to say whether this is due to selection biases (for example \citealt{Andrews2018} targeted the brightest discs), the low number statistics (even combined, there are only a handful of stars in these samples below 0.5 $M_\odot$) or if it is physical. 
Imaging these discs might seem harder because they are in general significantly fainter than around solar-mass stars, since the disc sub-mm flux strongly correlates with the host stellar mass \citep{Pascucci2016}. 
However, it should be kept in mind that the disc size also correlates with the stellar mass; in fact, the disc surface brightness is almost constant \citep{Tripathi2017,Andrews2018} across the disc population. 
Because interferometers like ALMA are sensitive to surface brightness, rather than absolute flux, the prospect to image discs around low mass stars looks encouraging. 
Future observations will thus provide the datasets necessary to test how the incidence of planets at 10s of AU in young systems depends on the mass of the central star.

\section{Conclusions}
\label{sec:conclusions}
In this paper we have investigated the planet gap opening mass, defined as relevant for ALMA continuum observations (i.e., in the dust, rather than in the gas), across stellar masses and for different distances from the star. We have highlighted how the dependence on the stellar mass is the net result of the competition between the two different effects: on one hand, gap opening depends on the planet-stellar mass ratio, favouring gap opening by lower mass planets around low mass stars. On the other hand, discs around low mass stars are geometrically thicker due to the reduced gravity, making gap opening more difficult due to the increased pressure forces.

We have shown that if we assume a dependence of stellar luminosity on stellar mass appropriate to the \citet{Siess2000} isochrones at $1$ Myr, the latter effect is more important than the former; in this case we would therefore predict that the gap opening mass \textit{decreases} with stellar mass and  that planet induced structure should therefore be more readily detectable in the case of more massive stars. For the \citet{Baraffe2015} tracks, the two effects almost exactly cancel each other; it is therefore a robust conclusion that there is no benefit in looking for planets around low mass stars. 
The gap opening mass also increases with the distance from the star, as expected in a flaring disc. 
We provide a simple scaling relation (see Eq. \ref{eqn:summary_Mmin} and figure \ref{fig:summary}) that expresses the gap opening mass as a function of orbital radius and stellar mass, where $A_{M_{*}}$ is the gap opening mass in Earth masses at a distance of $1$ au. This relation can readily be used in the interpretation of observations and is applicable at angular distances from the star that exceed the beam size. 

However the detailed interpretation of future imaging results needs to take into account the actual stellar
luminosities in the observed sample,  since the luminosity spread at a given mass introduces significant differences for individual discs. In general the stellar luminosity of each source will also be known and 
 we also provide relations to take  this into account when  estimating the gap opening mass, see Eq. \ref{eqn:Mmin_Ls_apl} and figure \ref{fig:MGOPM_Ls_apl}.
 
 Planets are often held responsible for the annular structures now ubiquitously observed in proto-planetary discs and future surveys will determine how the incidence of such structures depends on stellar mass. Our study has provided the framework within which the results of such surveys should be interpreted. 
 

\section*{Acknowledgements}
We thank an anonymous referee for their constructive criticism that significantly improved this paper. This work has been supported by the DISCSIM project, grant
agreement 341137 funded by the European Research Council under
ERC-2013-ADG and also by the European Union's Horizon
2020 research and innovation programme under the Marie
Sklodowska-Curie grant agreement No 823823 (DUSTBUSTERS). This work used the DIRAC Shared Memory Processing system at the University of Cambridge, operated by the COSMOS Project at the Department of Applied Mathematics and Theoretical Physics on behalf of the STFC DiRAC HPC Facility (www.dirac.ac.uk). This equipment was funded by BIS National E-infrastructure capital grant ST/J005673/1, STFC capital grant ST/H008586/1, and STFC DiRAC Operations grant ST/K00333X/1. DiRAC is part of the National E-Infrastructure. This work is part of the research programme VENI with project number 016.Veni.192.233, which is (partly) financed by the Dutch Research Council (NWO).



\bibliographystyle{mnras}
\bibliography{references} 




\appendix

\section{Dependence of the temperature on stellar luminosity}
\label{sec:append_temp}

It is common when estimating disc temperatures to assume the simple relation $T \propto L^{1/4}$. There are some refinements to make to this simplified version \citep{1997ApJ...490..368C,2001ApJ...560..957D}, that in practise conspire to make the relation between temperature and luminosity flatter. If the disc is optically thin to cooling by its own thermal radiation (as it is common in the outer parts of the disc), the temperature will also depend on the Planck mean opacity $\kappa_p$ and disc surface density. In addition, the relation should contain also the flaring angle $\phi \propto h/r$, setting how much of the stellar light is intercepted by the disc:
\begin{equation}
T \propto \left( \frac{h}{r} \frac{L}{\kappa_p \Sigma} \right)^{1/4}
\end{equation}

Let us exemplify the change introduced by these extra factors by first considering the variation in Planck mean opacity. At low temperatures a common behaviour is that $\kappa_p \propto T^2$ \citep[e.g.,][]{2003A&A...410..611S}. Using this fact, one gets that $T \propto L^{1/6} = L^{0.17}$, which is close to the result we get in section \ref{sec:spread}. The physical interpretation of this relation is that colder discs are less efficient at cooling due to a reduced Planck mean opacity.

Note that in section \ref{sec:spread} we hold $\Sigma$ constant and we vary only the stellar luminosity. It is worth noting instead that, when we vary the stellar mass in section \ref{sssec:dischr}, $\Sigma$ varies. It is a reasonable assumption that $\Sigma$ should increase with stellar mass, and this is the case for our models. This fact tends to flatten even further the temperature-luminosity relation when the stellar mass is varied. Finally, while we do not take this into account in this paper, the weaker gravity of low-mass stars (since $h/r \propto \sqrt{T/M}$) tends to flatten the relation even more, since discs around low mass stars are geometrically thicker and intercept more stellar light.


\bsp	
\label{lastpage}
\end{document}